\documentclass[10pt,letterpaper]{article}
\usepackage{opex3}
 \usepackage{graphicx,float}
 \usepackage{subfigure}
\usepackage{amsmath}
 \usepackage{dcolumn}
\usepackage{bm}
\usepackage{mathrsfs}
\usepackage{txfonts}
\usepackage{amssymb}
\usepackage[amssymb]{SIunits}
\usepackage{epsfig}
\usepackage{color}
\usepackage{cite}

\begin{document}

\title{Ultrabroadband nonreciprocal transverse energy flow of light in linear passive photonic circuits}

\author{Keyu Xia,$^{1,2,3,*}$ M. Alamri,$^{2}$ and M. Suhail Zubairy$^{1}$}  %

\address{$^1$Institute for Quantum Science and Engineering (IQSE) and Department of Physics and Astronomy, Texas A\&M University, College Station, Texas 77843-4242, USA\\
$^2$The National Center for Mathematics and Physics, KACST, P.O. Box 6086, \\ Riyadh 11442, Saudi Arabia\\
$^3$ARC Centre for Engineered Quantum Systems, Department of Physics and Astronomy,\\ Macquarie University, NSW 2109, Australia
}
\email{*keyu@physics.tamu.edu}


\begin{abstract}
Using a technique, analogous to coherent population trapping in an atomic system, we propose schemes to create transverse light propagation violating left-right symmetry in a photonic circuit consisting of three coupled waveguides. The frequency windows for the symmetry breaking of the left-right energy flow span over $80$~\nano \meter. Our proposed system only uses linear passive optical materials and is easy to integrate on a chip.
\end{abstract}


\bibliographystyle{osajnl}

\section{Introduction}

The integration of nonreciprocal photonic devices on Si or CMOS platforms has been challenging in the past decades.
The known optical nonreciprocity can be divided into two classes: forward-backward nonrecipricity (FBNR) and left-right nonrecipricity (LRNR). The first class of nonreciprocal component is well studied \cite{RPP67p717} and can be realized by various approaches using the magneto-optic effects \cite{fujita2000waveguide,haldane2008possible,zheng2009observation, OE20p18440,PRL107p173902,NaturePhotonOnchip}, nonlinearity \cite{Science335p447},  a modulation media \cite{NPhoton3p91,OptoAcousticIsolator1,OptoAcousticIsolator2}, optomechanics \cite{PRL102p213903,Optom2}, or magnetized plasmonic metal \cite{hadad2010magnetized,khanikaev2010one}. It can be used for optical isolators. 
To date, the LRNR, indicating the nonreciprocal light flow between left and right ports of photonic circuits,  has been discussed in an array of coupled waveguides by only a few research groups \cite{NPhys6p192,PRL103p093902,PRA82p043803}. If the complex optical potential causes the parity-time (PT) symmetry breaking, two coupled waveguides can show a LRNR light transfer in the transverse direction \cite{NPhys6p192,PRL103p093902}. The realization relies on the precise control of the active medium. Although not for optical isolators,  the photonic circuit with the left-right nonrecipricity may switch or route the incident light beams. 
%

The need for integration of optical nonreciprocal elements on a Si/CMOS chip platform is a long-standing problem. The realization of the nonreciprocal light propagation in a completely linear optical medium can strongly impact on both fundamental physics, and also vast applications for integrated optics because of the compatibility with the Si material and CMOS chips.
Again based on the PT symmetry breaking induced by a periodic modulation of complex optical potential, Feng et al. stated that they, for the first time, observed the nonreciprocal light propagation in a linear passive optical material \cite{Science333p729}. Unfortunately, they admitted their mistake \cite{Science333p38c} after Fan et al. commented on their work \cite{Science333p38b}. Another experimental realization of on-chip optical diodes using all-dielectric, passive, and linear silicon photonic crystal structures is reported by Wang et al. \cite{OE19p26948,SRep2p674}. They confidentially explained why it is possible to make optical diodes using a spatial symmetry breaking geometry in a passive and linear optical medium \cite{SRep2p674}.

In the LRNR, the input light is always localized in one waveguide \cite{NPhys6p192,PRL103p093902,PRA82p043803}. This is similar to the coherent population trapping in a three-level $\Lambda-$type atom \cite{CPT}. While the bending waveguide array can simulate well the quantum dynamics of atoms. The classical optical analogs of coherent population transfer \cite{ClassicCPTTheo1,ClassicCPTExp,LPR3p243} and population trapping in the continuum\cite{ClassicCPT} and Rabi oscillation \cite{ClassicRO} as well has been proved by Longhi's group. Although  the trapping of equal light in two waveguides has been discussed \cite{ClassicCPTTheo1}, the results did not show a valid LRNR of light flow. A theory work indicated that the critical large nonlinearity is necessary to induce nonrecipricity \cite{PRA82p043803} in two evanescently coupled waveguides. However the LRNR in the transverse light flow has been recently observed \cite{NPhys6p192,PRL103p093902}. Moreover, the three/many-body systems behavior essentially different from the simple two-body system studied in \cite{PRA82p043803}. At least, the optical trapping in coupled three waveguides analogous to the atomic CPT can not be achieved in an optical system composing of two waveguides. It is interesting if one can realize the nonreciprocal wave propagation in evanescently coupled linear and passive waveguides. We expect to achieve the LRNR in a waveguide array. This is the motivation of our work.

Here we propose simple methods to generate the second class of optical nonreciprocity in an array of three coupled waveguides only making from linear, passive optical materials. We focus on the nonreciprocal transverse energy flow between left and right optical waveguides. This left-right nonreciprocity does not violate the Lorentz reciprocity theorem. Thanks to a small dispersion of a linear waveguide, our system can behave in a nonreciprocal manner in an ultrabroad band. The results by numerical simulation of beam propagating method (BPM) and solving the coupled mode equation (CME) demonstrate the breaking of symmetry of transverse energy flow.  

\section{Setup and model}

\begin{figure}
\centering
 \includegraphics[width=0.6\linewidth]{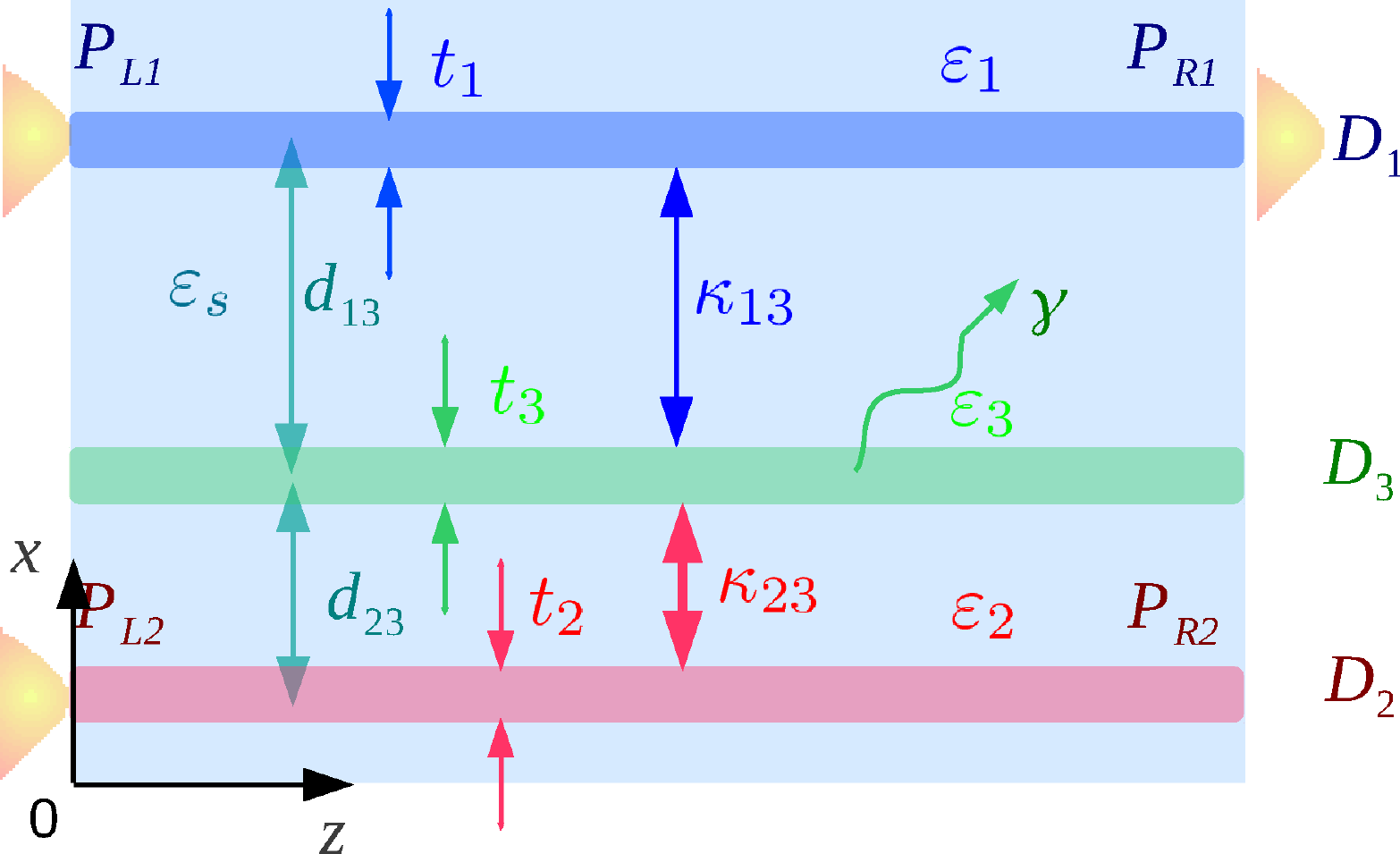}\\
\caption{Light trapping in a photonic circuit consisting of three waveguides embedded in a $W$ wide, $L$ long substrate. The coupling $\kappa_{13} \ll \kappa_{23}$. $d_{13,23}$ is the distance between two waveguides. The dielectric constants of substrate is $\varepsilon_s$, while $\varepsilon_{i}$ ($i\in \{1,2,3\}$) is the constant profile of corresponding individual waveguides D$_i$ without others. The widths of waveguides are $t_i$, respectively.}\label{fig:system}
\end{figure}

Our system, shown in Fig.~\ref{fig:system}, is composed of three coupled waveguides embedded in a wafer of width $W$ and length $L$. The middle waveguide D$_3$ couples to the waveguides D$_1$ and D$_2$. We assume that the coupling between the waveguides D$_1$ and D$_2$ is negligible. We also assume that the two side waveguides are lossless but some loss can be included in the middle one. In our photonic system, eigenmodes in the individual waveguide exchange energy via their evanescent fields when two waveguides are close. The couplings are denoted as $\kappa_{13}$ and $\kappa_{23}$, and decrease as the distance $d_{13,23}$ between two waveguides increases. The coupling between D$_1$ and D$_2$ is assumed vanishing because these two waveguides are far enough from each other. The field in D$_3$ decays exponentially with a constant $\gamma$ that can be controlled \cite{TuneLoss1,TuneLoss2,NPhys6p192,Science333p729,PRL103p093902}.

The evolution of power of field in photonic circuits can be studied either by numerical solving the Helmholtz equation or with the derived coupled mode theory. However the later presents a clearer physic understanding. We first present the approach based on Helmholz equation.

\subsection{Beam propagating method}
We consider our system to be two-dimensional (2D). This is reasonable if the size of waveguide in the $y$ direction is much larger than that in the $x$ direction. The propagation of the field $E$ in the photonic circuits in a 2D space can be described by the Helmholtz equation, which takes the form 
\begin{equation} \label{eq:HelmholtzE}
\frac{\partial^2 E}{\partial z^2}+ \frac{\partial^2 E}{\partial x^2}+ k_0^2 \varepsilon (x,z)E=0\,,
\end{equation}
where $z$ is the propagating direction, $x$ is the transverse direction, and $k_0=\frac{2\pi}{\lambda}$ is the wave vector of field with wavelength $\lambda$ in the free space. The dielectric constant $\varepsilon(x,z)$ plays the role of the optical potential. In Eq.~\ref{eq:HelmholtzE}, we assume that $\frac{\partial E}{\partial t}\sim 0$ and it is reasonable for our interest in the steady-state behavior of the system. We can resolve the field $E$ into its slowly varying amplitude $\psi$ and a fast oscillating factor, $E=\psi e^{\pm j\beta z}$, where the propagation constant $\beta=k_0 n_{eff}$. $n_{eff}$ is the effective index of waveguides. The sign before $\beta$ indicates the propagating direction: minus (plus) for the propagation along the positive (negative) $z$ direction. In the paraxial approximation $\left|2j\beta \frac{\partial \psi}{\partial z} \right| \gg \left|\frac{\partial^2\psi}{\partial z^2} \right|$, the amplitude of the field in the photonic circuit evolves according to the equation \cite{CMT1}

\begin{equation} \label{eq:AE}
 \mp 2j\beta\frac{\partial \psi}{\partial z} \approx \frac{\partial^2\psi}{\partial x^2}
+k_0^2 \left[\varepsilon (x,z)-n^2_{eff}\right]\psi \,.
\end{equation}

We numerically solve Eq.~(\ref{eq:AE}) with BPM to simulate the propagation of field with a spatial resolution $\delta z=1$~\micro \meter~ and $\delta x=0.1$~\micro \meter. A finer spatial grid gives the same results. 
Throughout simulation, we use the zero-order eigenmode profile $\psi(x,x_0,z=0,L)$ of TE mode by solving the eigen equation \cite{CMT1}, where $x_0=x_0^{(i)}$ ($i\in\{1,2\}$) is the center position of input ports $P_{L1,L2,R1}$. This mode profile is very close to a Gaussian function $\exp\left(-\left(x-x_0 \right)^2/2 w_{\text{p}}^2\right)$ with a half width of waist $w_{\text{p}}=1$~\micro \meter~.
The profiles $\psi(x,x_0^{(1)}, 0)$ and $\psi(x,x_0^{(1)},L)$ corresponds to the input field launching into the waveguide D$_1$ from the port $P_{L1}$ at $z=0$ and $P_{R1}$ at $z=L$, respectively. While the profile $\psi(x,x_0^{(2)}, 0)$ means an input to the port $P_{L2}$ at $z=0$. The intensity of the field at the peak is unity. This profile is very close to the fundamental eigenmode of D$_1$ or D$_2$. 

Our simulation focuses on a light with wavelength $\lambda_0=1.55$~\micro\meter~, which is of interest in optical communications. The photonic circuit can be integrated in a wafer with a substrate dielectric constant $\varepsilon_s=10.56\varepsilon_0$, where $\varepsilon_0$ is the permittivity of free space. We consider weakly guiding waveguides with $\varepsilon_{\text{core}}=10.76\varepsilon_0$ in order to ensure the valid of our BPM and CME method. The imaginary part of the dielectric constant in the middle waveguide D$_3$ is $\Im[\varepsilon_3]=-0.01\varepsilon_0$. This induces a loss of $\gamma=11.6$~\milli \meter$^{-1}$ according to our numerical simulation. The propagation constant is calculated by solving the eigenvalue equation \cite{CMT1} of TE mode. 

In our numerical method, the Neumann boundary condition (NBC) is used to greatly suppress the reflection field from the transverse boundary. A small reflection, which can be a practical noise from the boundary of device in experiments, is responsible for the background noise of our numerical results.

\subsection{Coupled mode equation method}
Before discussing the results, we present the equivalent, but physically transparent, coupled mode equation approach to explain our system.

The light field $E$ in photonic circuits can be expressed as a supermode of eigenmodes $E_i$ of individual waveguide D$_i$, i.e.,
\begin{equation}
 E =  \sum_{i=1,2,3}A_iE_i= \sum_{i=1,2,3} A_i \psi_i e^{\pm j\beta_i z}\,,
\end{equation}
where the amplitude of eigenmode in the $i$th waveguide is denoted by $A_i$ ($i\in \{1, 2, 3\}$). Here $\psi_i$ is the slowly varying envelope of eigenmode $E_i$ and $\beta_i$ is the corresponding propagation constant, which can be controlled by designing the dielectric constant $\varepsilon_i$ of individual waveguide D$_i$ and its width $t_i$. This parameter also depends on the wavelength of light. According to the coupled mode theory \cite{CMT1,CMT2,CMT3} derived from the Helmholtz equation, the dynamics of the modal amplitudes is described by

\begin{subequations}\label{eq:CME}
 \begin{align}
  \frac{\partial A_1}{\partial z} &= j\Delta_{13} A_1+ j \kappa_{13}(z) A_3\\
  \frac{\partial A_2}{\partial z} &= j\Delta_{23} A_1+j \kappa_{23}(z) A_3\\
  \frac{\partial A_3}{\partial z} &= j \kappa_{13}(z) A_1 + j \kappa_{23}(z) A_2 -\gamma A_3\,,
 \end{align}
\end{subequations}

where $\Delta_{13}=\beta_1-\beta_3$ and  $\Delta_{23}=\beta_2-\beta_3$ are the phase mismatch between waveguides D$_1$ (D$_2)$ and D$_3$. The power in the $i$th waveguide is evaluated by $P_i=A_i A_i^*$. The total power in the system is $P=\sum_i P_i$. The phase mismatching and coupling in the CME Eq.~(\ref{eq:CME}) can be derived from Helmholtz equation Eq.~(\ref{eq:HelmholtzE}).
These parameters are dependent on $k_0$ and the optical potential $\varepsilon(x,z)$. The coupling is given by \cite{CMT1,CMT2,CMT3}
\begin{equation}\label{eq:coupling}
\kappa_{mn}=\frac{k_0^2\int_s \delta\varepsilon_n  E_m^*E_n dS}{\beta_m\int_s  E_m^*E_m dS} \,,
\end{equation} 
where $s$ denotes the cross section of space and $\delta \varepsilon_n=\varepsilon(x,z)-\varepsilon_n$ at the cut position $z$. 
For simplicity's sake, we have assumed weakly guiding waveguides and the relation $\kappa_{mn}=\kappa_{nm}$. We also neglect the second-order spatial derivatives of the amplitude $A_i$ and the small self phase shifts due to the perturbation of neighbor waveguides. A full study of the relation of parameter to the Helmholtz equation has been presented by Hardy et al. \cite{CMT2,CMT3}. Note that the coupled mode theory presents a general model. In contrast, the numerical results depend on the structure of system, and only provides one of many implementations. The different structures can lead to the same set of parameters in the CME. On the other hand, the numerical simulation presents a full picture of light in photonic circuits.

Next we turn to our idea about how to create optical nonreciprocity in the transversal energy flow in three coupled waveguides by giving a connection of our classic photonic circuit to a quantum system. 

Due to the equivalence between the Helmholtz equation in photonic circuits and the Schr\"{o}dinger equation in quantum mechanics, the behavior of light propagating in a photonic circuit is similar to the dynamics of the internal atomic states of a quantum system \cite{ClassicCPTTheo1,ClassicCPTExp,LPR3p243}. For example, the normalized light power trapped in optical waveguides plays the role of atomic population. Optical nonreciprocities in an array of coupled waveguides can then be considered as the trapping of input light on demand. As is well known, in a $\Lambda-$type three-level atomic systems, we can adiabatically create a target state independent of the initial state of system via the so-called coherent population trapping (CPT) \cite{CPT}. Our system is analogous to such a $\Lambda$-type three-level system. Just as in CPT in the atomic system, we expect to trap most light energy in a selected optical waveguide by suitably controlling the coupling between the waveguides. This is the basis of  the optical nonreciprocity studied in this paper.
Note that the LRNR we propose here is substantially different from the FBNR used for optical isolator on the basis of the breaking of Lorentz reciprocity theorem \cite{Haus, RPP67p717}. In the former case, both two sources input into ports in the left hand side but their responses come out from the right hand side, while the source and the response must exchange in the later. As a result, $\int {\bf J}_s^{(1)} \cdot {\bf E}_{R}^{(2)} d{\bf S} = \int {\bf J}_s^{(2)} \cdot {\bf E}_{R}^{(1)} d{\bf S} =0$ for our 2D case, where the response ${\bf E}_{R}^{(i)}$ at port $P_{R1}$ is the electric field created by a source ${\bf J}_s^{(j)}$ at port $P_{L1,L2}$ with $i\neq j$ and $i,j\in \{1,2\}$, because the source and response are separated in space, i.e. ${\bf J}_s^{(1)} \cdot {\bf E}_{R}^{(2)} =  {\bf J}_s^{(2)} \cdot {\bf E}_{R}^{(1)} =0$.  Thus the implementation of the LRNR in a linear, passive medium does not violate the Lorentz reciprocity theorem \cite{Haus, RPP67p717}, which requires the exchange of the place of source and response.

\subsection{Connection between two methods}
\begin{figure}
\begin{center}
 \includegraphics[width=0.44\linewidth]{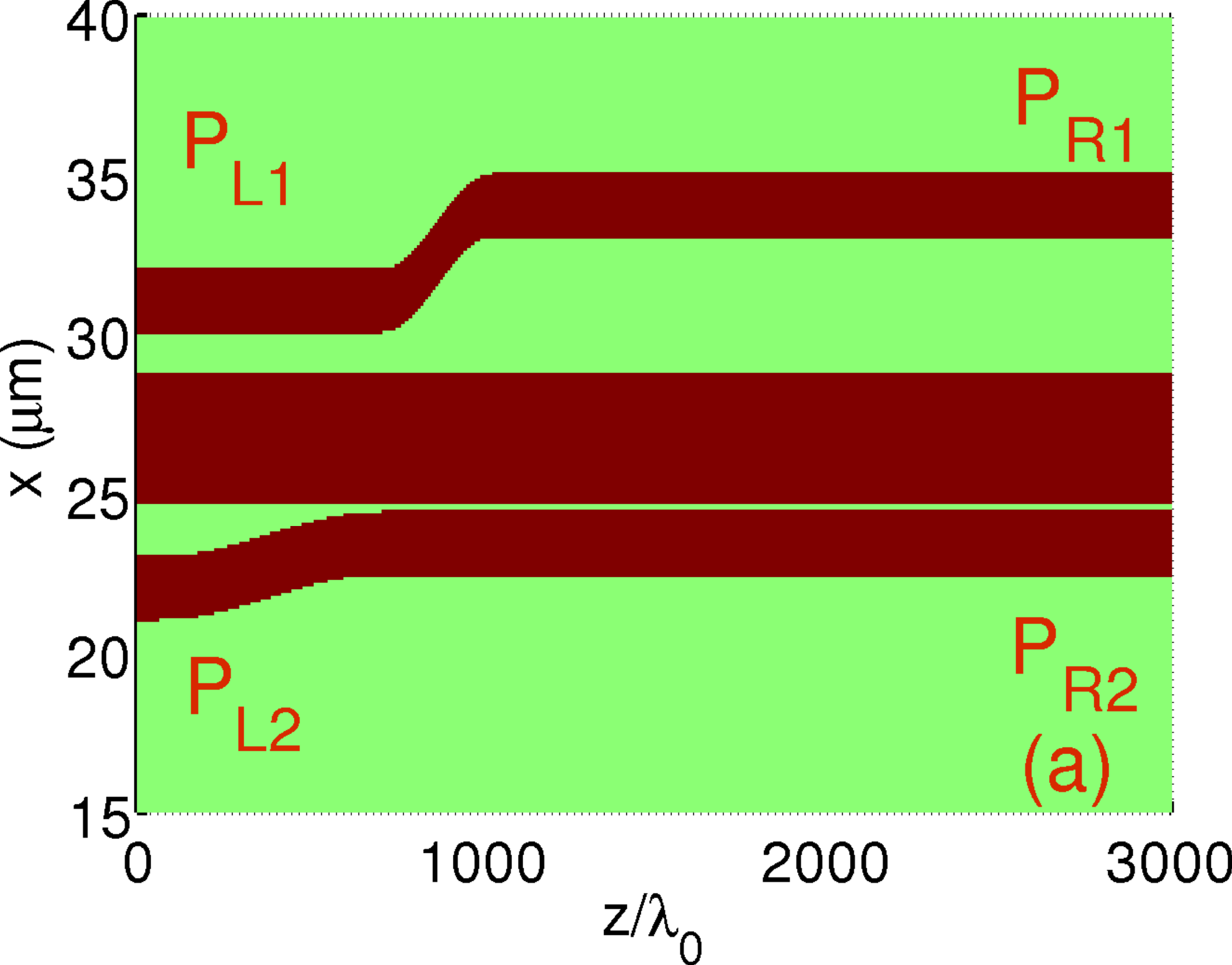}
 \includegraphics[width=0.53\linewidth]{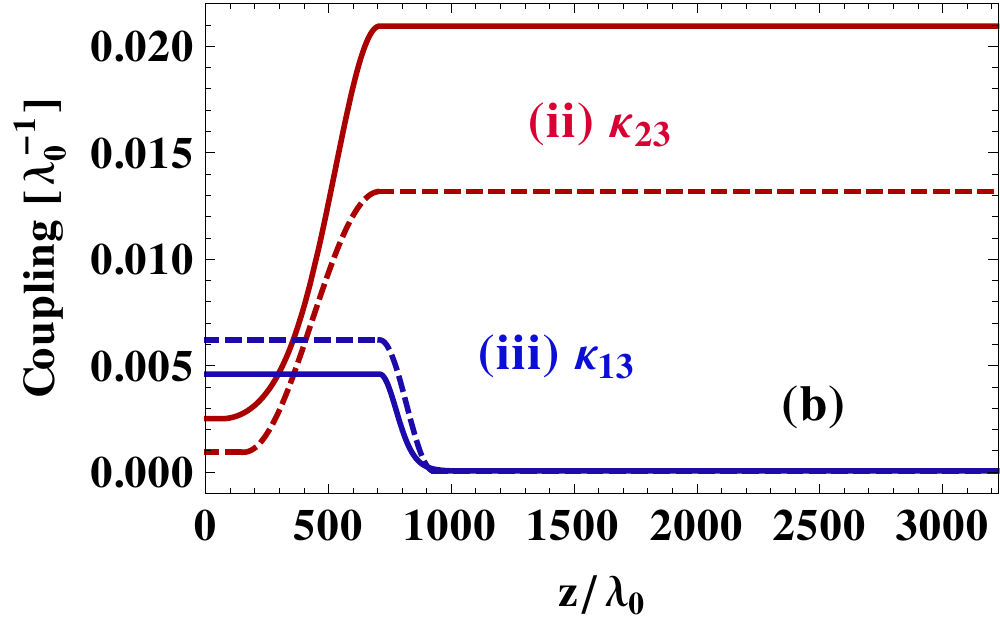}
\caption{(a) The waveguide structure for left-right nonrecipricity.
The straight waveguide D$_3$ is $4$~\micro\meter~ wide around its center $w_3(0)=26.7$~\micro\meter. The waveguides D$_1$ and D$_2$ with width $t_1=t_2=2$~\micro\meter~ are curves along their varying central positions $w_1(z)$ and $w_2(z)$ defined as $w_1[z][\micro\meter]=31$ for $z<1.1$~\milli\meter; $w_1[z][\micro\meter]=34$ for $z>1.6~\milli\meter$ and $w_1[z][\micro\meter]=31+3(1+\sin(2\pi(z-1350)/1000))/2$ for $1.1~ \milli\meter \leq z \leq 1.6~ \milli\meter$. $w_2[z]$ is constant $22~\micro\meter$ for $z < 0.1~\milli\meter$ and $23.4~\micro\meter$ for  $z > 1.1~\milli\meter$. During the transition region, $w_2[z][\micro\meter]=22 + 1.4(1+\sin(2\pi(z-600)/2000))/2$ for $0.1~\milli\meter \leq z \leq 1.1 \milli\meter$.
 (b) The coupling as a function of propagating distance $z$. Blue lines (i) for $\kappa_{13}$, red lines (ii) for $\kappa_{23}$. Solid lines for coupling rates are evaluated by Eq.~\ref{eq:coupling}, while dashed lines indicates coupling rates for fitting the numerical results. Detailedly, the coupling rates for fitting are
$\kappa_{13}(z)$[\milli\meter$^{-1}$]=$4$ for $z<1.1$~\milli\meter; $\kappa_{13}(z)$[\milli\meter$^{-1}$]=$0.03$ for $z>1.44$~\milli\meter~ and $\kappa_{13}(z)$[\milli\meter$^{-1}$]=$0.03+3.97(1-\sin(2\pi(z-1270)/680))/2.0$ for $1.1$~\milli\meter $\leq z \leq 1.44$~\milli\meter. While $\kappa_{23}(z)$ is $0.6$~\milli\meter$^{-1}$ for $z<0.24$~\milli\meter~ and $8.5$~\milli\meter$^{-1}$ for $z>1.1$~\milli\meter, and $0.6+ 7.9(1+\sin(2\pi(z-670)/1720))/2.0$ for $0.24$~\milli\meter $\leq z \leq 1.1$~\milli\meter. Here $\lambda_0=1.55~\micro\meter$. }\label{fig:coupling}
\end{center}
\end{figure}

The structure of photonic circuit to create the nonreciprocal transverse energy flow of light is shown in Fig.~\ref{fig:coupling} (a). Through our system, we assume no loss in waveguide $D_1$ and $D_2$. The dielectric constant is $\varepsilon_{\text{s}}=10.56 \varepsilon_0$ in the substrate, while it is $\varepsilon_{\text{core}}=10.76 \varepsilon_0$ in core of $D_1$ and $D_2$. In waveguide $D_3$, the dielectric constant in core is $\varepsilon_3=(10.76-0.01 i)\varepsilon_0$. 
%
%
To fit the numerical results, the coupling $\kappa_{13}$ and $\kappa_{23}$ are assumed to vary corresponding to the central positions $w_{1}(z),w_{2}(z)$ and $w_{3}(z)$ of waveguides. The other parameters for CME are given by:
\begin{align}
 \gamma= &11.6~ \milli \meter^{-1} \,, \\
 \Delta_{13}= &\Delta_{23}=-23 ~\milli \meter^{-1} \,. 
\end{align}
The mismatch of propagating constant $\Delta_{13}= \Delta_{23}=-23$~\milli \meter$^{-1}$ is obtained by solving the eigenvalue of zero-order TE mode.
To simulate the varying gaps between waveguides, we assume two gradient changing coupling strength $\kappa_{13}$ and $\kappa_{23}$
in the propagating direction for Eq.~\ref{eq:CME} as shown in Fig.~\ref{fig:coupling} (b).
Two waveguides in the same chip always couples to each other even if the coupling strength is very small. To consider this coupling, we assume small values as the distance between two waveguides are large. The intensity in waveguides given by the CMEs change very slightly if we neglect this small coupling. 
We note that, in the absence of loss in the waveguide $w_3$, the system is reciprocal (not shown here). However if we include loss in the middle waveguide, we create left-right nonreciprocities. 

Let us assume that a light with unit amplitude is incident on the port $P_{L1}$ or $P_{L2}$ at $z=0$.
If the photonic circuit is reciprocal, the transmission from port $P_{R1}$ or $P_{R2}$ exchanges as well if the incident exchange.
However, in the case of left-right nonrecipricity, the light launched into port $P_{L1}$ and $P_{L2}$ always effectively transfer to the waveguide D$_1$ and comes out from port $P_{R1}$. 
If we use constant couplings $\kappa_{13}$ and $\kappa_{23}$, the LRNR is obtained but the transmissions are small. The energy trapped in waveguide D$_1$ also decays because part of the energy couples to the middle waveguide from which the energy is lost into the environment at a rate $\gamma$. To avoid a strong coupling of energy between D$_1$ and D$_3$, we gradually change the distance between the two side waveguides and the middle one to guarantee an adiabatic process. In addition, large phase mismatchings $\Delta_{13}$ and $\Delta_{23}$ are used to suppress the energy coupling to waveguide D$_3$. In the output side, we decouple the waveguides D$_1$ and D$_3$ by introducing a large distance to keep the light energy in D$_1$ almost constant. The profile of the mode is also kept stable after $z/\lambda_0=1500$. 
The couplings used in the CME for fitting the following numerical results are dashed lines shown in Fig.~\ref{fig:coupling} (b). The solid lines are numerically evaluated by Eq.~\ref{eq:coupling}. These coupling are strongly dependent on the distance $d_{13,23}$. In spite the exact numerical solution of coupling rates $\kappa_{13}$ and $\kappa_{23}$ from Eq.~\ref{eq:coupling} is different from the numbers we use to fit the distribution of fields below, it provide us a good guide for the fitting function.

\section{Results}

\begin{figure}
\begin{center} 
 \includegraphics[width=0.45\linewidth]{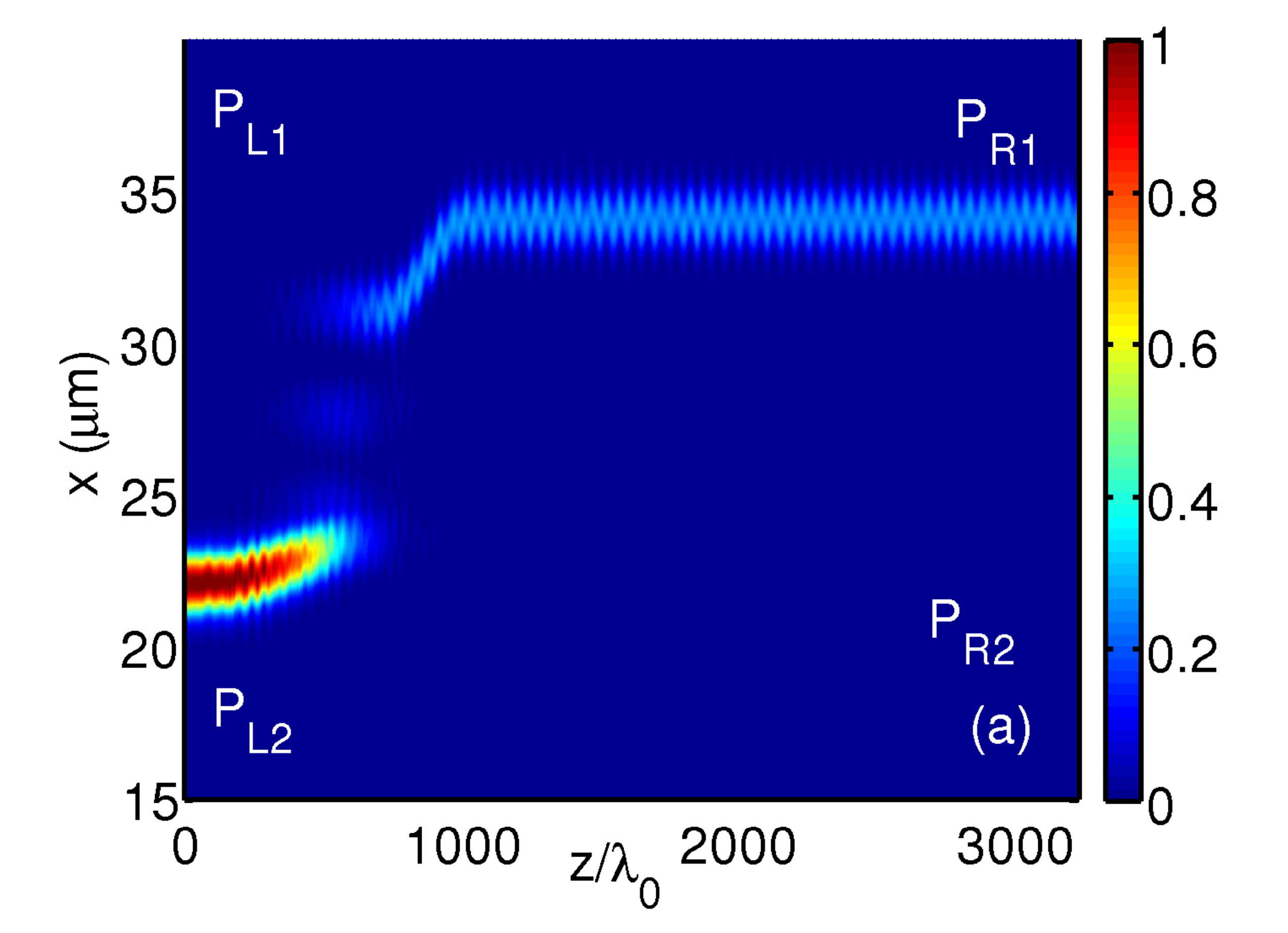}
\includegraphics[width=0.45\linewidth]{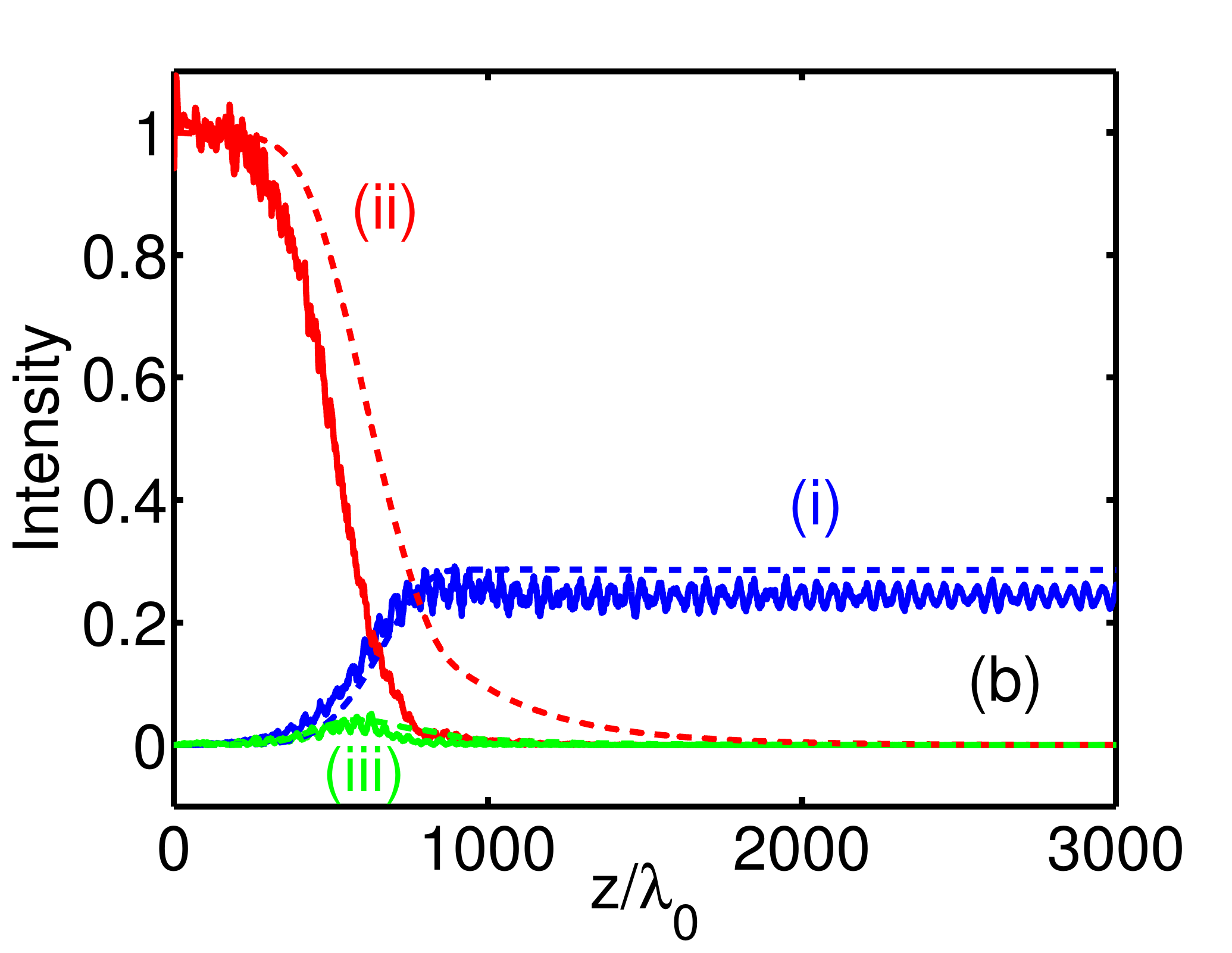}\\
 \includegraphics[width=0.45\linewidth]{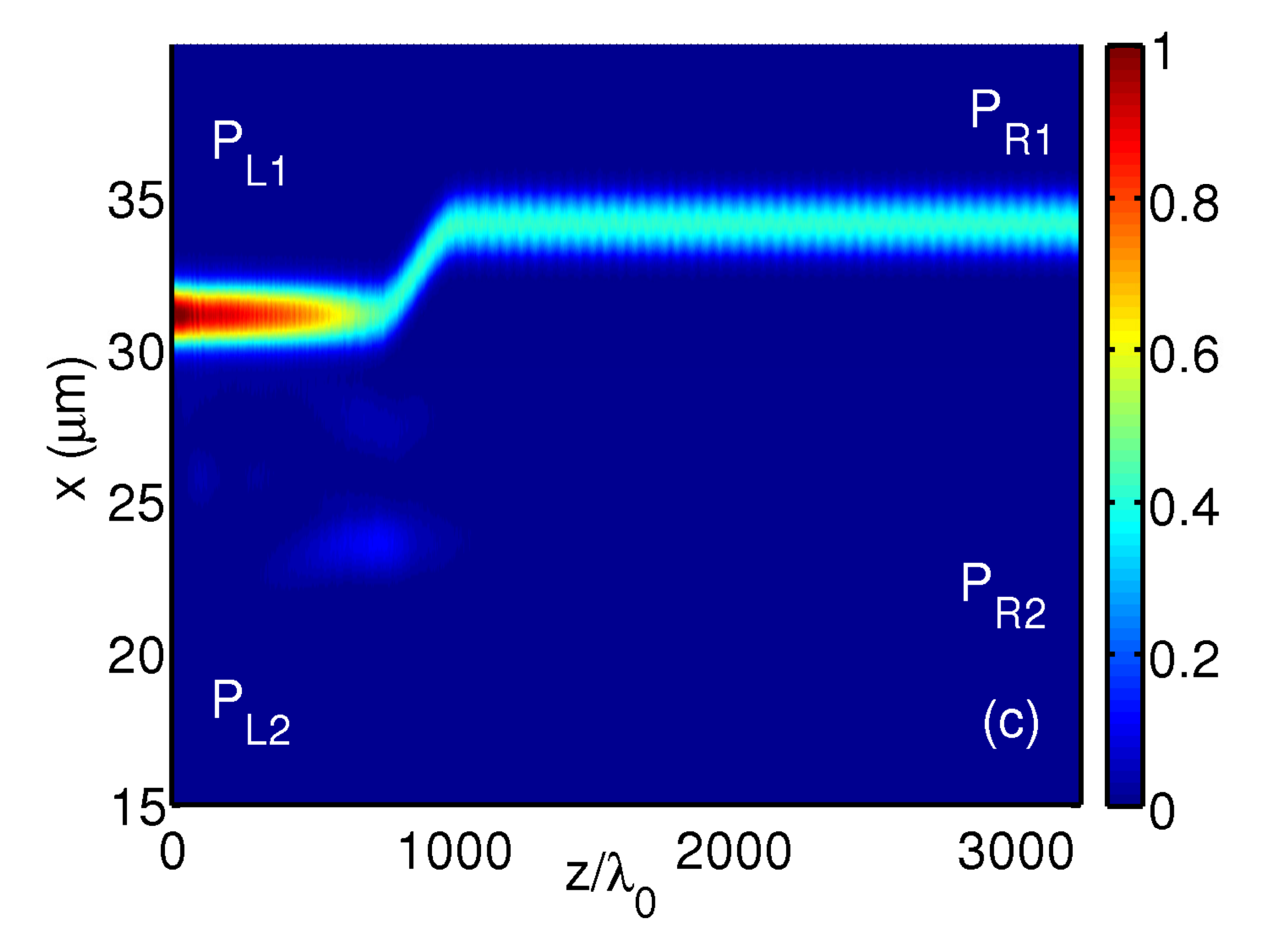}
\includegraphics[width=0.45\linewidth]{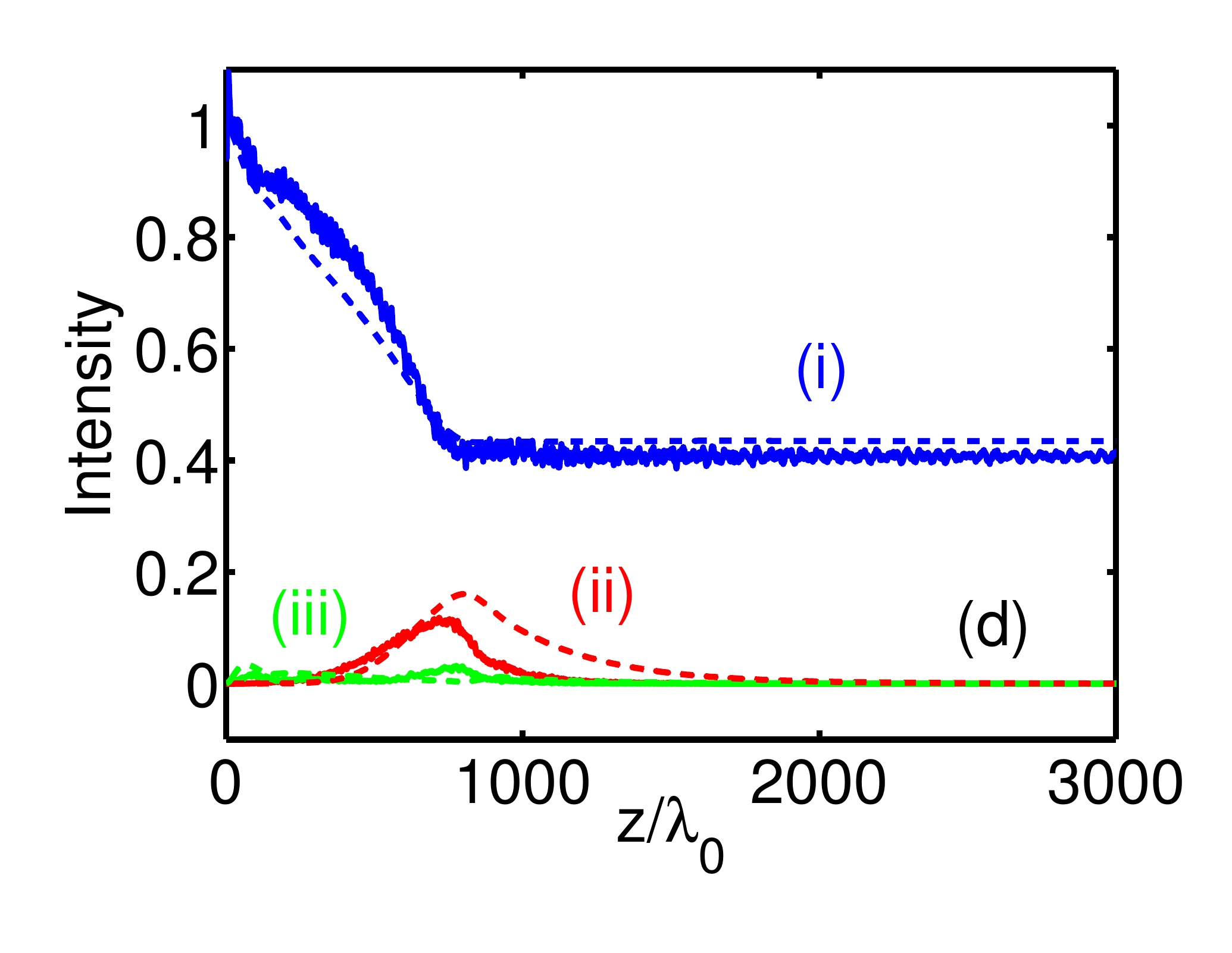}\\
\caption{\label{fig:NLRNon} Left-right nonreciprocity corresponding to Fig.~\ref{fig:coupling} (a). The field propagates from left to right in photonic circuits. (a) Light incident into the waveguide D$_2$;  (c) light enters the waveguide D$_1$; (b) and (d) Intensities of field at the middle of waveguides D$_1$ [blue lines (i)],  D$_2$ [red lines (ii)] and D$_3$ [green lines (iii)]. Dashed thin lines are the corresponding plots by solving Eq.~(\ref{eq:CME}).}
\end{center}

\end{figure}

Now we study the left-right nonreciprocity where we have nonreciprocal light transfer in the transverse direction \cite{NPhys6p192,PRL100p103904}. Similar to coherent population trapping in quantum optics, we can trap most light energy in the selected waveguide D$_1$ by designing a weak coupling $\kappa_{13}$ in comparison with $\kappa_{23}$. Our numerical results shown in Figs.~\ref{fig:NLRNon} (a) and (c) demonstrate a left-right nonreciprocal transverse energy flow. Whatever port $P_{L1}$ or $P_{L2}$ we choose to lauch the light into, most of the light is trapped in the waveguide D$_1$, and comes out from the same port $P_{R1}$ (blue lines). The transmission for light input into port $P_{L2}$ is about $25\%$ (Figs.~\ref{fig:NLRNon}(b)) but it increases to $40\%$ if the light is incident into port $P_{L1}$ (Figs.~\ref{fig:NLRNon}(d)). The light in the waveguide D$_2$ leaks to D$_3$ and subsequently is absorbed as it propagates. The contrast ratios of light intensities in waveguides D$_1$ and D$_2$ are higher than $29$~dB in both cases.

So far we presented results based on the numerical solution of the Helmholtz equation. Next we compare these results with the solution of the CME Eq.~(\ref{eq:CME}), see dashed lines in Figs.~ \ref{fig:NLRNon}(b) and~ \ref{fig:NLRNon}(d). Clearly, the coupled mode theory agrees well with the numerical simulation in detail. The intensity of light in waveguide D$_1$ in Figs. \ref{fig:NLRNon}(b) are slightly higher than the numerical results of BPM method.  A small discrepancy is that the light in waveguide D$_2$ decays slower than the numerical results mainly because the loss in waveguide D$_2$ in the bending region is not included in the coupled mode theory Eq.~(\ref{eq:CME}). A full coupled mode theory \cite{CMT2,CMT3} involving higher modes and many parameters corresponding to the structure of system may present a better fitting of numerical results by the BPM. However, to provide a clearer physical understanding of the LRNR behavior we use the simple model, Eq. ~(\ref{eq:CME}). In spite of small discrepancy, the coupled mode theory still fits numerical results detailedly.

\begin{figure}
 \begin{center}
  \includegraphics[width=0.45\linewidth]{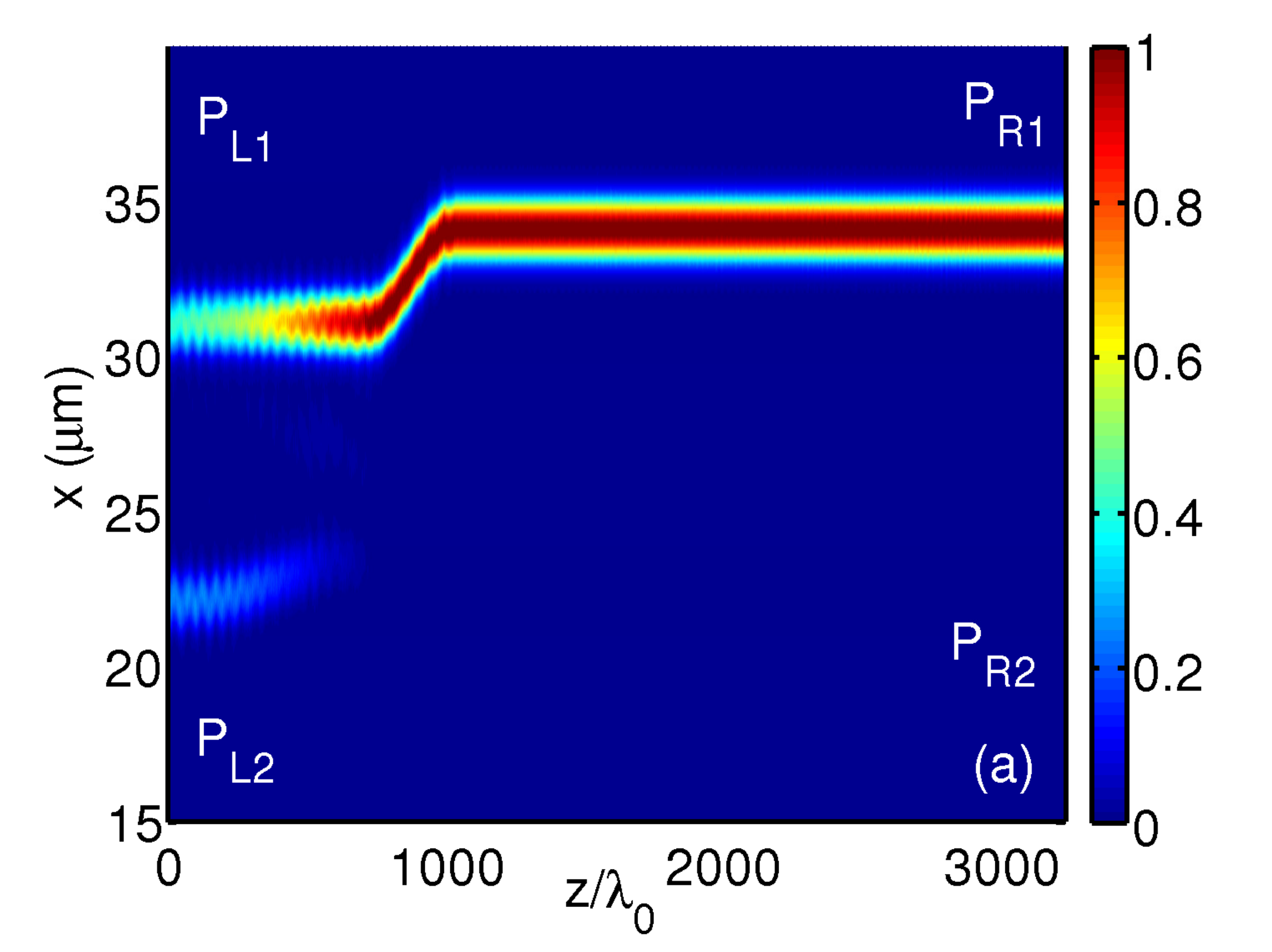}
  \includegraphics[width=0.45\linewidth]{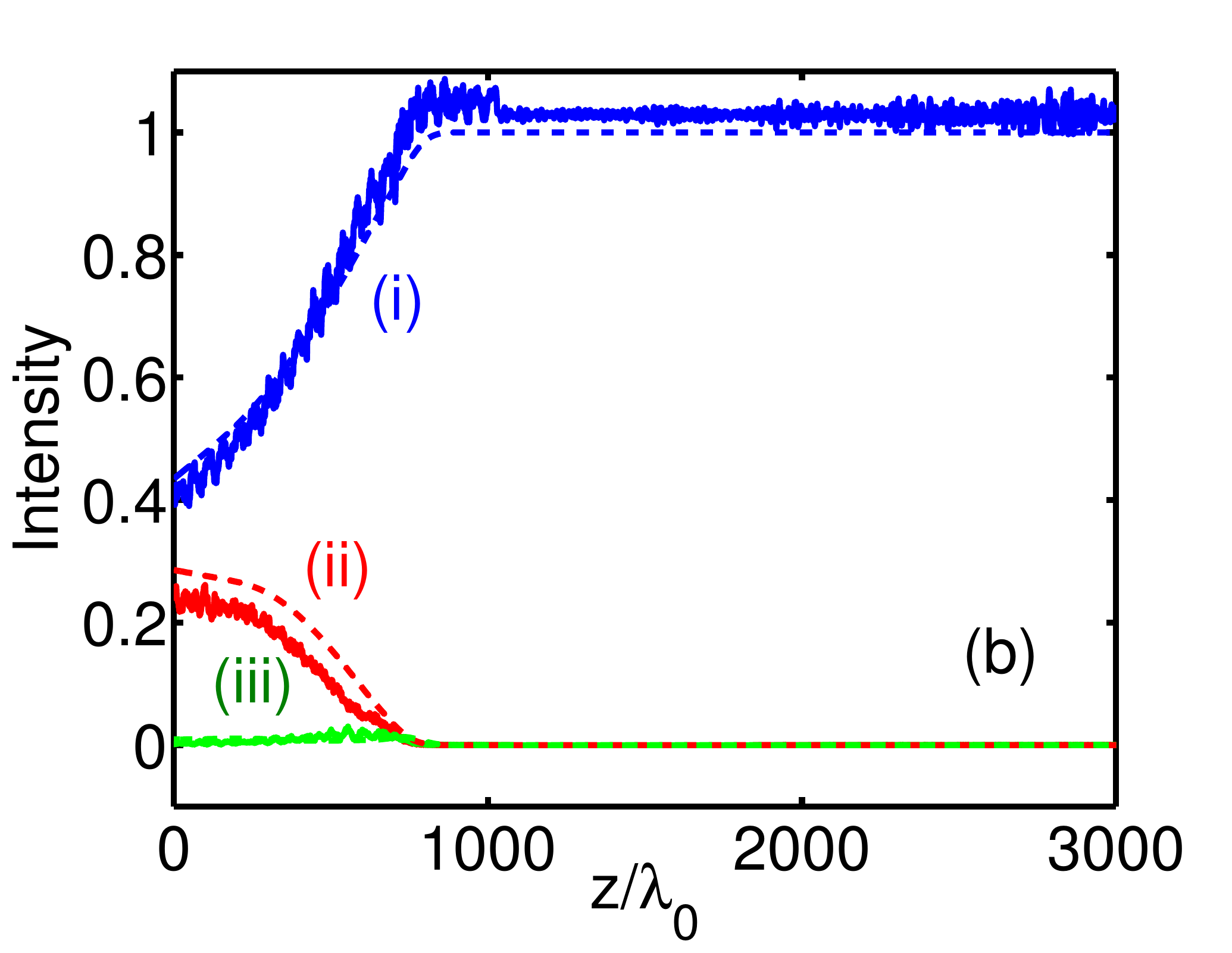}\\
\caption{Numerical simulation of propagation of light in Fig.~\ref{fig:coupling}(a). The field incident into port P$_{R1}$ propagates from right to left. (a) Distribution of  field (intensity), (b) Intensities of field at the middle of waveguides D$_1$ [blue lines (i)], D$_2$ [red lines (ii)],D$_3$ [green lines (iii)]. Dashed lines are the fitting plots evaluated by Eq.~(\ref{eq:CME}) using the same parameters as in Fig.~\ref{fig:NLRNon}.}\label{fig:FBR}
 \end{center}
\end{figure}

It is interesting to check whether the device displays the FBNR, a counterpart of LRNR, because the former is the basis for optical isolators. 
According to the Lorentz reciprocity theorem \cite{Haus} and Fan et al. \cite{Science333p38b}, the FBNR is impossible in a linear, time-independent medium. Our numerical simulation agrees with it and demonstrates the forward-backward reciprocity, as shown in Fig.~\ref{fig:FBR}. 
To check the forward-backward reciprocity, we interchange the source and the response and solve Eqs. (\ref{eq:AE}) and (\ref{eq:CME}). The light is incident into the port P$_{R1}$ and the output from ports P$_{L1}$ and P$_{L2}$ are monitored. Again, the results by solving Eq. (\ref{eq:CME}) using the same parameter and coupling rates fit the numerical results by BPM well. As predicted by the Lorentz Reciprocity theorem, the transmission from P$_{R1}$ to P$_{L1}$ (P$_{L2}$) in the numerical simulations equals to those from P$_{L1}$ (P$_{L2}$) to P$_{R1}$. So the Lorentz Reciprocity theorem still rules the dynamics of our system.

\begin{figure}
\begin{center}
\includegraphics[width=0.45\linewidth]{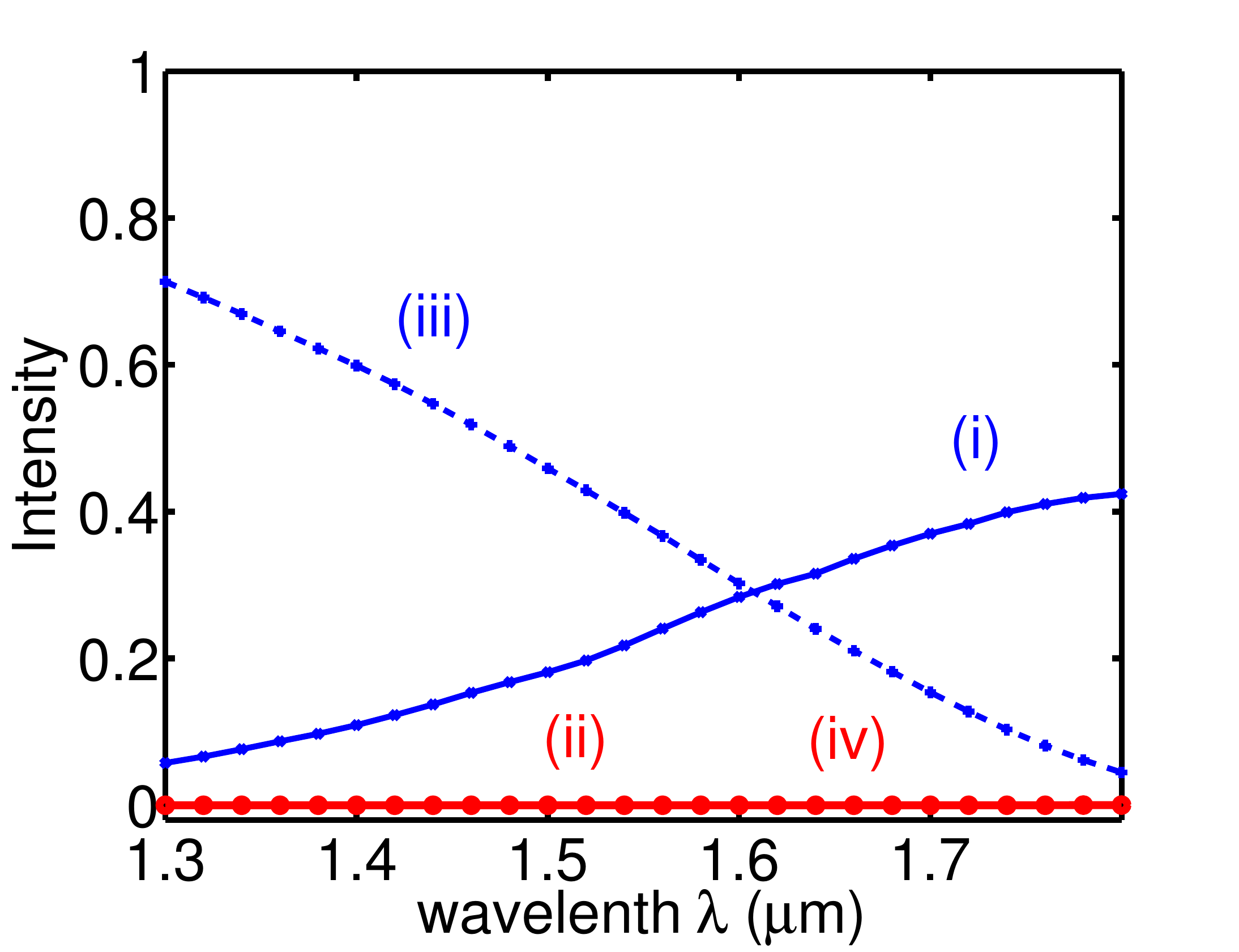}\\
\caption{\label{fig:NonrepFreq} Nonreciprocal transmission as a function of frequency of input light using structure as in Fig.~\ref{fig:coupling}(a). Thin blue lines (i) and (iii) show the light trapped in waveguide D$_1$, thick red lines (ii) and (iv) show the light energy in D$_2$. Solid lines for light launching to D$_2$, dashed lines for light input into D$_1$.}
\end{center}
\end{figure}

The bandwidth in which the propagating light show nonreciprocal behavior is an important feature. We study the frequency dependence of the transmission spectra in Fig.~\ref{fig:NonrepFreq} by numerical simulations.  We numerically calculate the transmissions using the same Gaussian profile for input in BPM method but solve the eigen equation \cite{CMT1} for the propagation constant for different wavelength. As the widths and refractive indices of the two side waveguides D$_1$ and D$_2$ are the same, the propagation constants $\beta_1$ and $\beta_2$ are equal. Thus the phase mismatch $\Delta_{13}$ is equal to $\Delta_{23}$ ideally. However each propagation constant itself and the coupling rates are dependent on the wavelength of input field. As a result, the transmission to P$_{R1}$ from P$_{L1}$ (P$_{L2}$) decreases (increases) gradually as the wavelength of incident light increases. Our system traps more than $24\%$ in waveguide D$_1$ from $1.56$~\micro \meter~ to $1.64$~\micro \meter. So it has an ultrabroadband nonreciprocal window over $80$~\nano\meter. In the nonreciprocal windows, the light in waveguide D$_2$ is always vanishing because it couples to the lossy channel D$_3$. Next we concentrate our discussion in the nonreciprocal window of interest. It can be clearly seen in Fig.~\ref{fig:NonrepFreq}, whatever waveguide the light is incident to, more than $24\%$ energy is trapped in D$_1$ and comes out of port $P_{R1}$.
The frequency-dependence of transmission comes from the change of eigenmode profiles, propagation constants and their couplings $\kappa_{13}$ and $\kappa_{23}$, which are also dependent on the profiles of eigenmodes and wavelength [ref. to Eq.~(\ref{eq:coupling})]. The deviation in fabrication may result in a small difference between $\Delta_{13}$ and $\Delta_{23}$. However the transmission change slightly if $\Delta_{13} \approx \Delta_{23}$. A longer bending waveguide can tune the coupling between waveguides slower but is not necessary to provide a wider nonrecipricity window because it also changes the effective coupling length and the propagation constants are dependent on the wavelength as well.

\begin{figure}
\begin{center}
\includegraphics[width=0.3\linewidth]{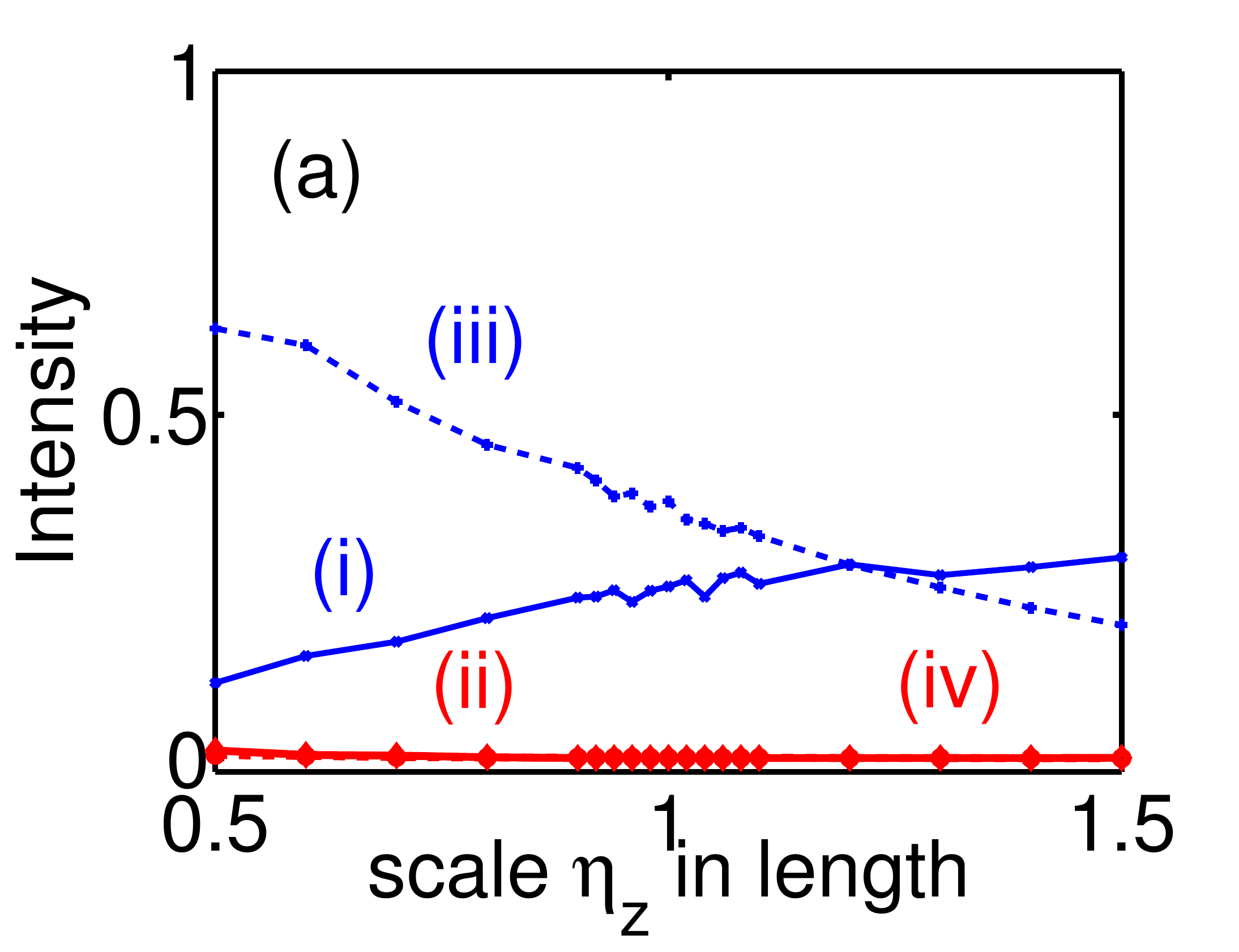}
\includegraphics[width=0.3\linewidth]{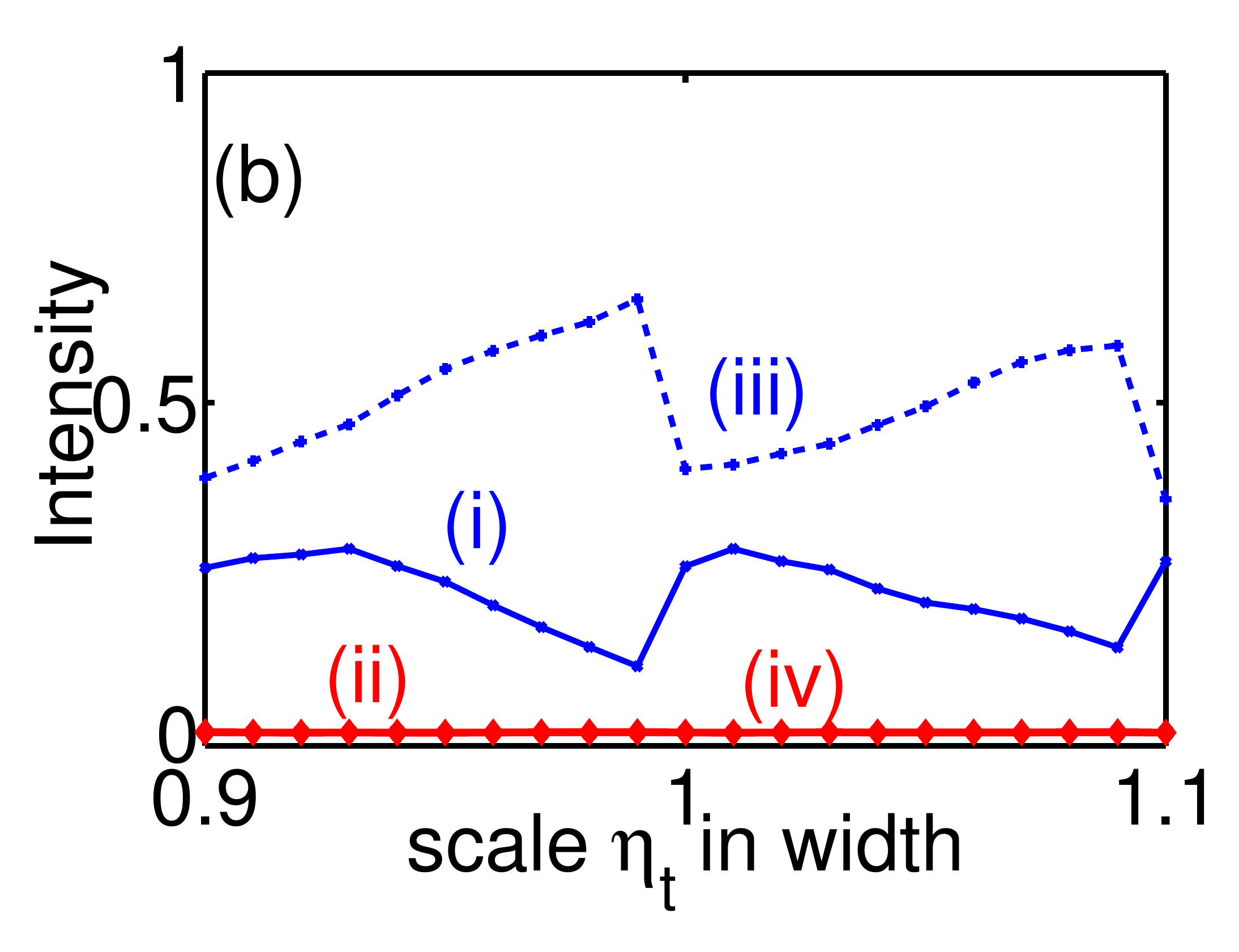}
\includegraphics[width=0.3\linewidth]{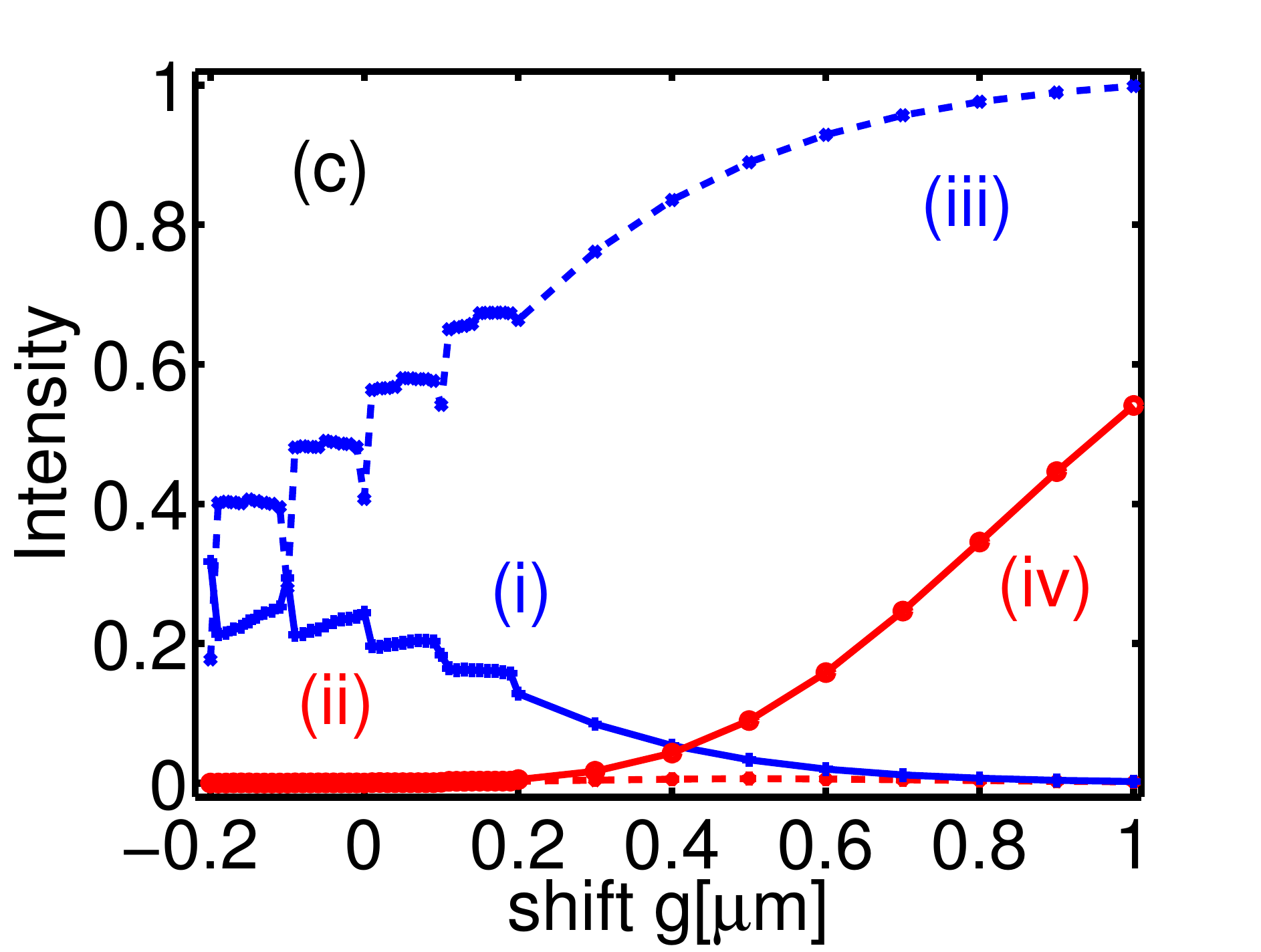}
\caption{\label{fig:geom} Nonreciprocal transmission in the structure as in Fig.~\ref{fig:coupling}(a) changes in the length (a), width (b) of, gap (c) between waveguide. (a) scale the device by $\eta_z$ in the z-direction; (b) scale the width of waveguides D$_1$ and D$_2$; (c) shift the center of waveguides by $g$ as $w_1[z] + g$ and $w_2[z] - g$. Thin blue lines (i) and (iii) show the light trapped in waveguide D$_1$, thick red lines (ii) and (iv) show the light energy in D$_2$. Solid lines for light launching to D$_2$, dashed lines for light input into D$_1$.}
\end{center}
\end{figure}

For a practical application, the performance of device need be robust against small deviation in structure. Figure~\ref{fig:geom} shows how robust the nonreciprocal transfer of light is when the length, width of and gaps between waveguides change. 
It can be seen from Fig.~\ref{fig:geom}(a) that the nonreciprocal performance varies slowly as the total length $\eta_z L$ of device changes. There is more than $24\%$ of input light is trapped in D$_1$ when the device scales in the z-direction from $\eta_z=1.0$ to $\eta_z=1.3$. As shown in Fig.~\ref{fig:geom}(b), the light trapped in D$_1$ oscillates as a function of the width $\eta_t t_1 =\eta_t t_2$ but it is stable for $1.0 \leqslant \eta_t \leqslant 1.02$, which means that the width of waveguides can vary $40$~\nano\meter. In contrast, the LRNR of our design is more sensitive to the distance between waveguides. The transmission is larger than $21\%$ if the shift/offset of waveguides $g$ is negative. It means a smaller distance between waveguides is preferable. While the nonreciprocity deteriorates rapidly as the distance increases.   

\begin{figure}
\begin{center}
\includegraphics[width=0.3\linewidth]{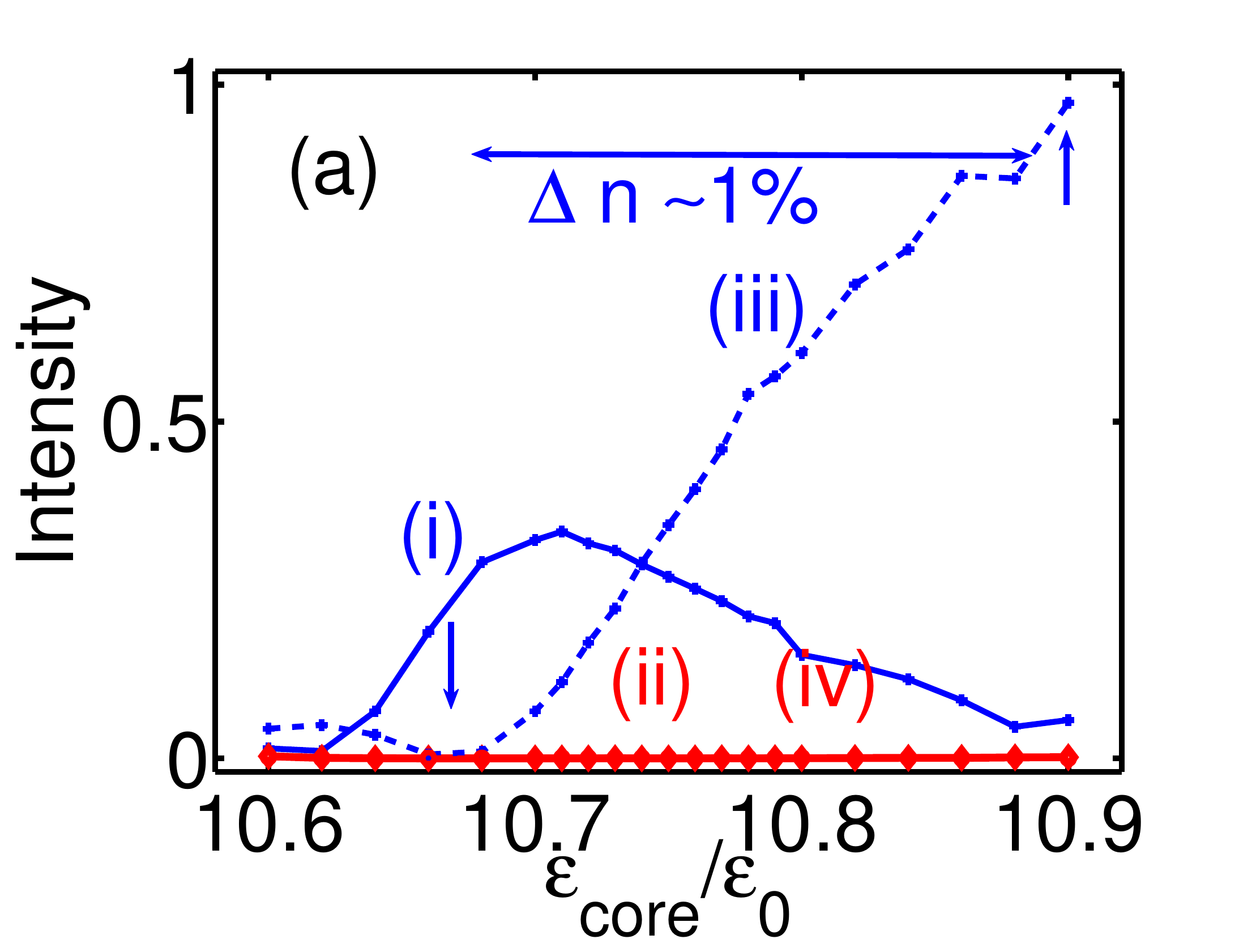}
\includegraphics[width=0.3\linewidth]{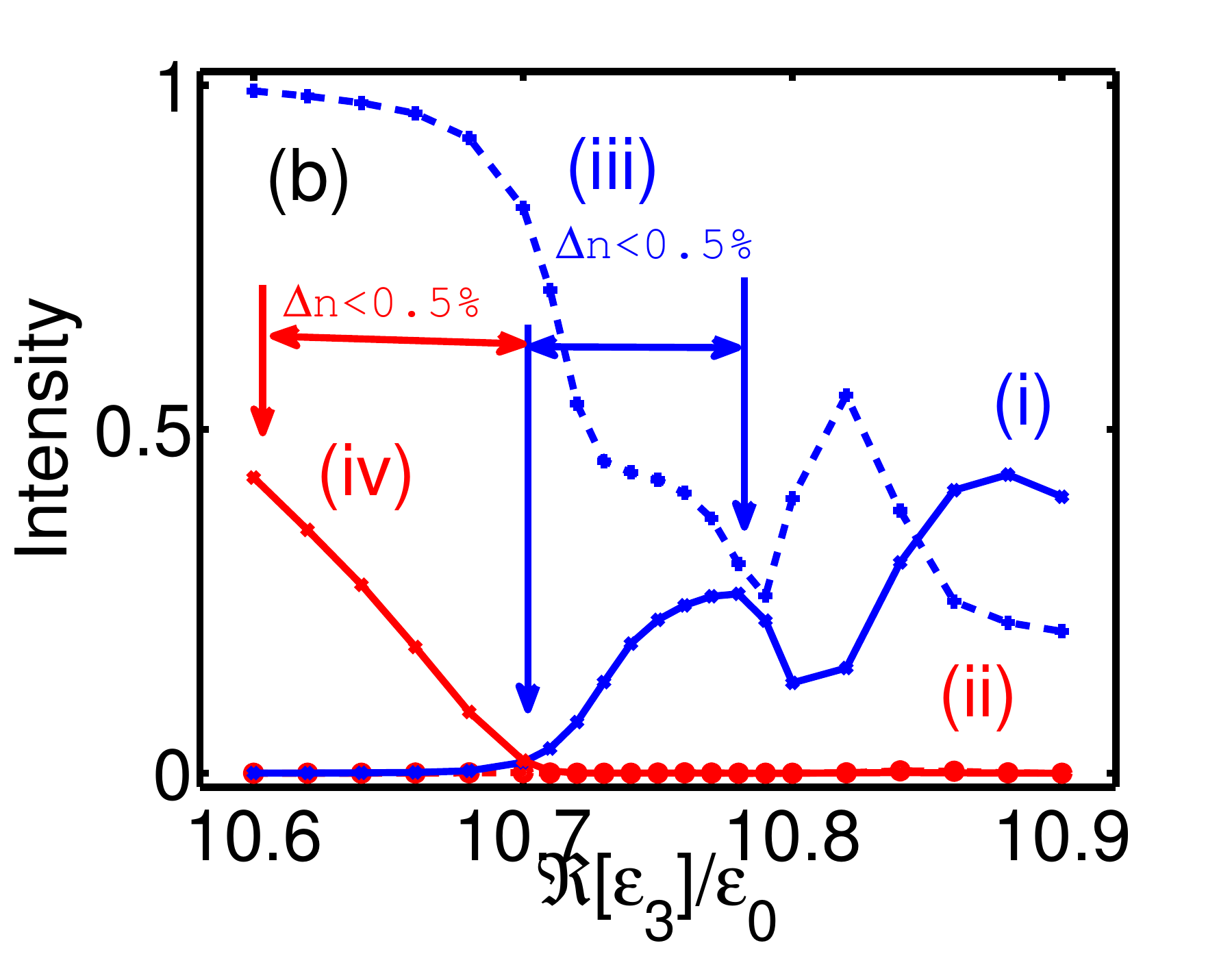}
\includegraphics[width=0.3\linewidth]{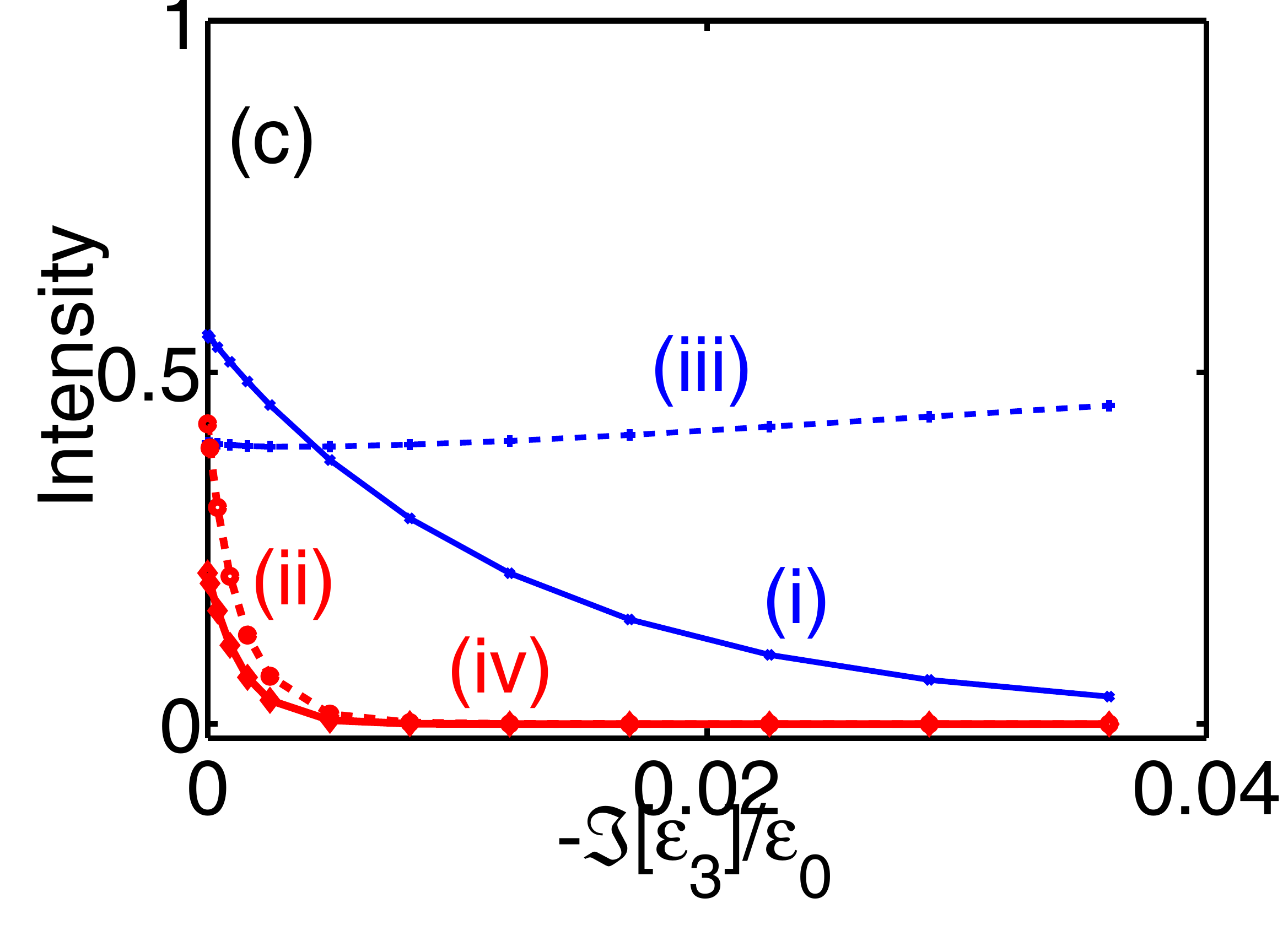}
\caption{\label{fig:DEC} Nonreciprocal transmission as the structure as in Fig.~\ref{fig:coupling}(a) changes of the global refractive index $\Re[\varepsilon_{core}]$ (a), refractive index $\Re[\varepsilon_3]$ (b) and loss $\Im[\varepsilon_{core}]$ of waveguide D$_3$. Thin blue lines (i) and (iii) show the light trapped in waveguide D$_1$, thick red lines (ii) and (iv) show the light energy in D$_2$. Solid lines for light launching to D$_2$, dashed lines for light input into D$_1$.}
\end{center}
\end{figure}

The deviation of dielectric constant in fabrication changes the mistmatching of propagation constants and the coupling rate as well. First, we check the performance when the global dielectric constant $\varepsilon_{core}$ changes in all waveguides, as shown in Fig.~\ref{fig:DEC}(a). The dielectric constant $\varepsilon_{core}$ need be accurately engineered to pursue for a good LRNR behavior. Only the region $10.73 \leqslant\varepsilon_{core} / \varepsilon_0 \leqslant 10.77$ is useful to trap light in waveguide D$_1$. When $\varepsilon_{core} / \varepsilon_0$ changes from $10.68$ to $10.9$ corresponding to $\Delta n/n_{core} \sim 1\%$, the output from P$_{R1}$ is switched from ``on'' (``off'') to `` off'' (``on'') for input to P$_{L2}$ (P$_{L1}$). Then the output is investigated as the dielectric constant $\varepsilon_3$ of waveguide D$_3$ changes only. For $\Re[\varepsilon_3]/ \varepsilon_0<10.7$, no LRNR displays in our system. When $\Re[\varepsilon_3]$ is larger, the LRNR occurs and the light trapped in D$_1$ fluctuates as the refractive index increases. However, more than $25\%$ of light can be trapped in D$_1$ over the region of $10.76\leqslant \Re[\varepsilon_3]/ \varepsilon_0 \leqslant 10.78$. In contrast, the LRNR is very robust against the loss of waveguide D$_3$. The light in D$_2$ decreases rapidly as the loss increases. The light trapped in D$_1$ is stable for an input to P$_{L1}$ and decays exponentially for an input to P$_{L2}$. When $\Im[\varepsilon_3]/ \varepsilon_0 < -0.005$ corresponding to $\gamma \geqslant 60$~\centi\meter$^{-1}$, no light in D$_2$ and there is only light in D$_1$. In the range of $-0.01\leqslant \Im[\varepsilon_3] / \varepsilon_0\leqslant -0.005$, the system can trap more than $25\%$ of light in D$_1$.

Our simulations show a relative flexible parameters to the LRNR. The existing modern technology can fabricate the device in a much more accuracy. Therefore a linear, passive medium can display highly optical nonrecipricity in our design.

\begin{figure}
\begin{center}
\includegraphics[width=0.45\linewidth]{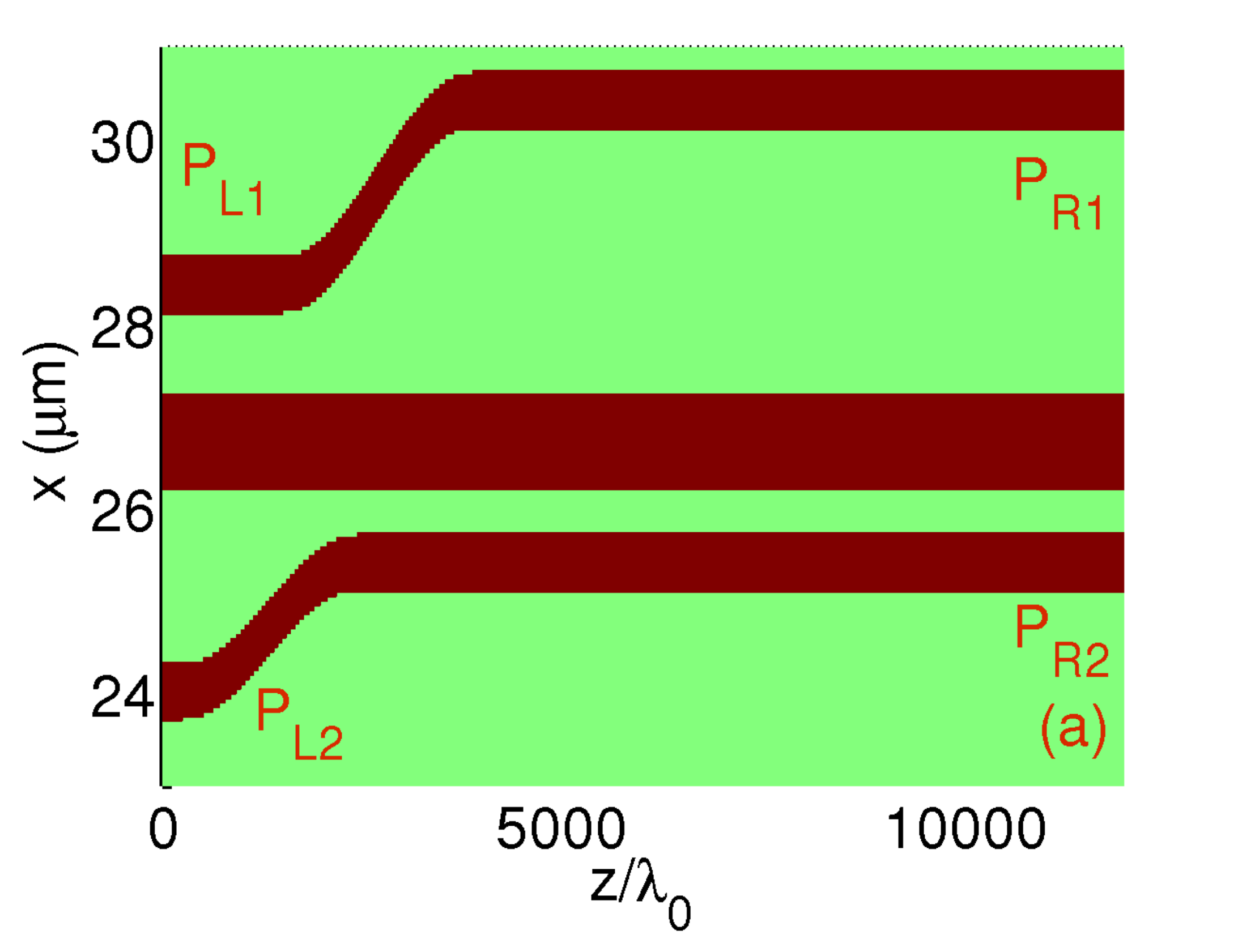}
\includegraphics[width=0.45\linewidth]{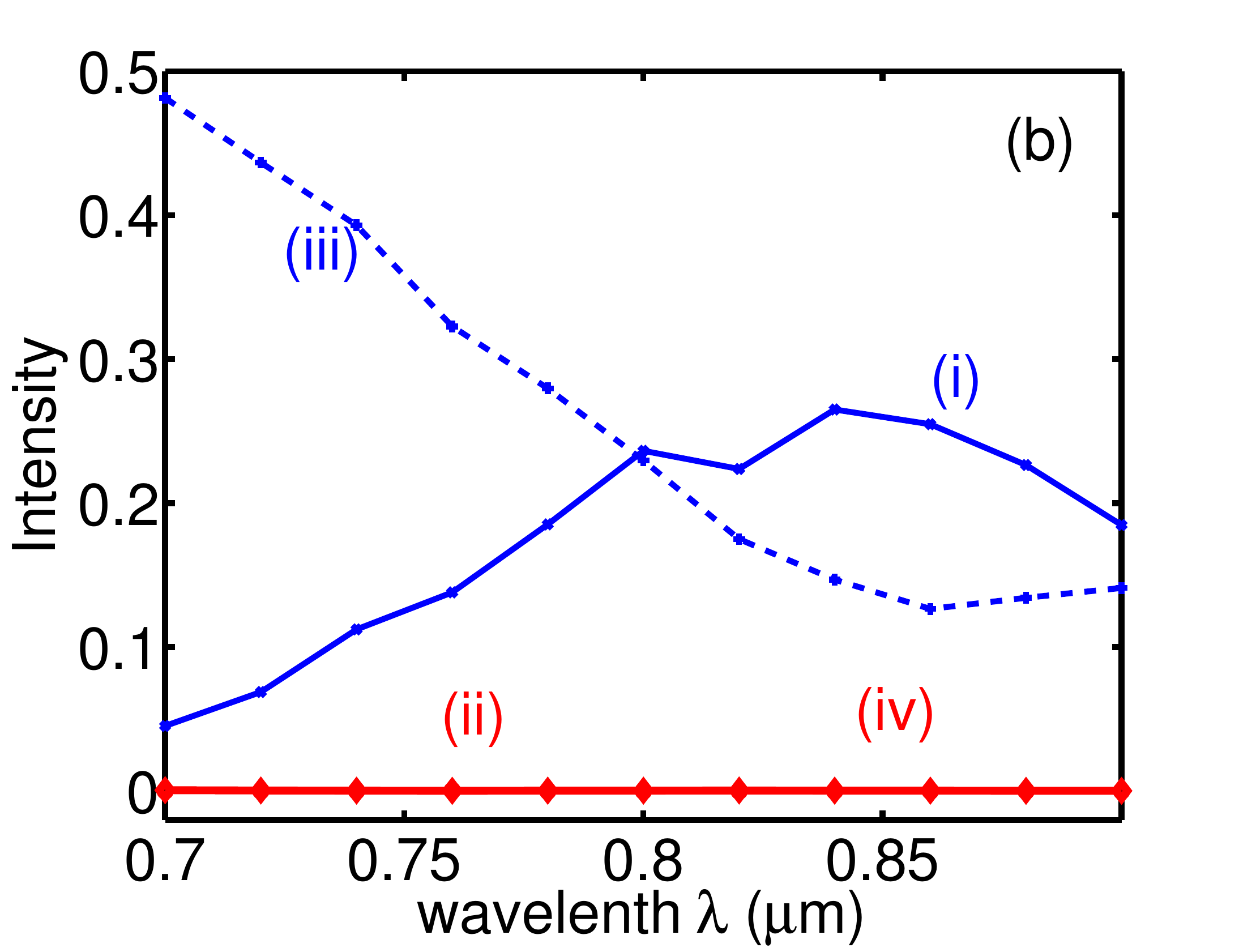}\\
\caption{\label{fig:800nm} (a) The waveguide structure for left-right nonreciprocity at $\lambda=800$~\nano\meter. 
Similar to Fig.~\ref{fig:system}(a),
the straight waveguide D$_3$ is $1$~\micro\meter~ wide around its center $w_3(0)=26.7$~\micro\meter. The waveguides D$_1$ and D$_2$ with width $t_1=t_2=0.6$~\micro\meter~ are curves along their varying central positions $w_1(z)$ and $w_2(z)$ defined as $w_1[z][\micro\meter]=28.4$ for $z<1.2$~\milli\meter; $w_1[z][\micro\meter]=30.4$ for $z>3.1~\milli\meter$ and $w_1[z][\micro\meter]=28.4+2(1+\sin(2\pi(z-2150)/3800))/2$ for $1.2~ \milli\meter \leq z \leq 3.1~ \milli\meter$. $w_2[z]$ is constant $24~\micro\meter$ for $z < 0.2~\milli\meter$ and $23.4~\micro\meter$ for  $z > 1.95~\milli\meter$. During the transition region, $w_2[z][\micro\meter]=24 + 1.4(1+\sin(2\pi(z-1075)/3500))/2$ for $0.2~\milli\meter \leq z \leq 1.95 \milli\meter$. 
(b) Nonreciprocal transmission as a function of frequency of input light in (a). Thin blue lines (i) and (iii) show the light trapped in waveguide D$_1$, thick red lines (ii) and (iv) show the light energy in D$_2$. Solid lines for light launching to D$_2$, dashed lines for light input into D$_1$. The Gaussian profile of input light is adjusted to be $w_p=0.375$~\nano\meter wide.}
\end{center}
\end{figure}

\begin{figure}
\begin{center} 
 \includegraphics[width=0.45\linewidth]{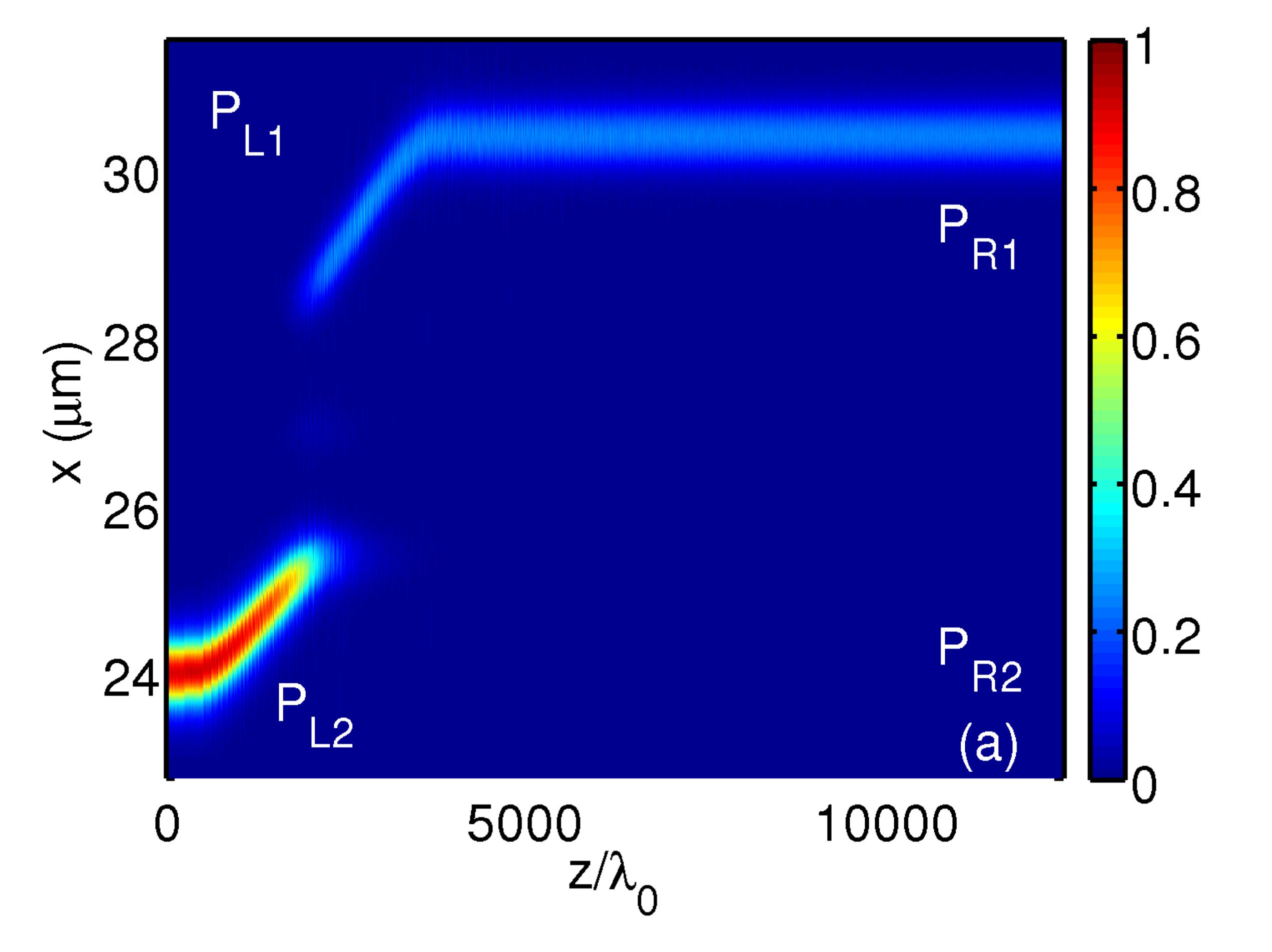}
\includegraphics[width=0.45\linewidth]{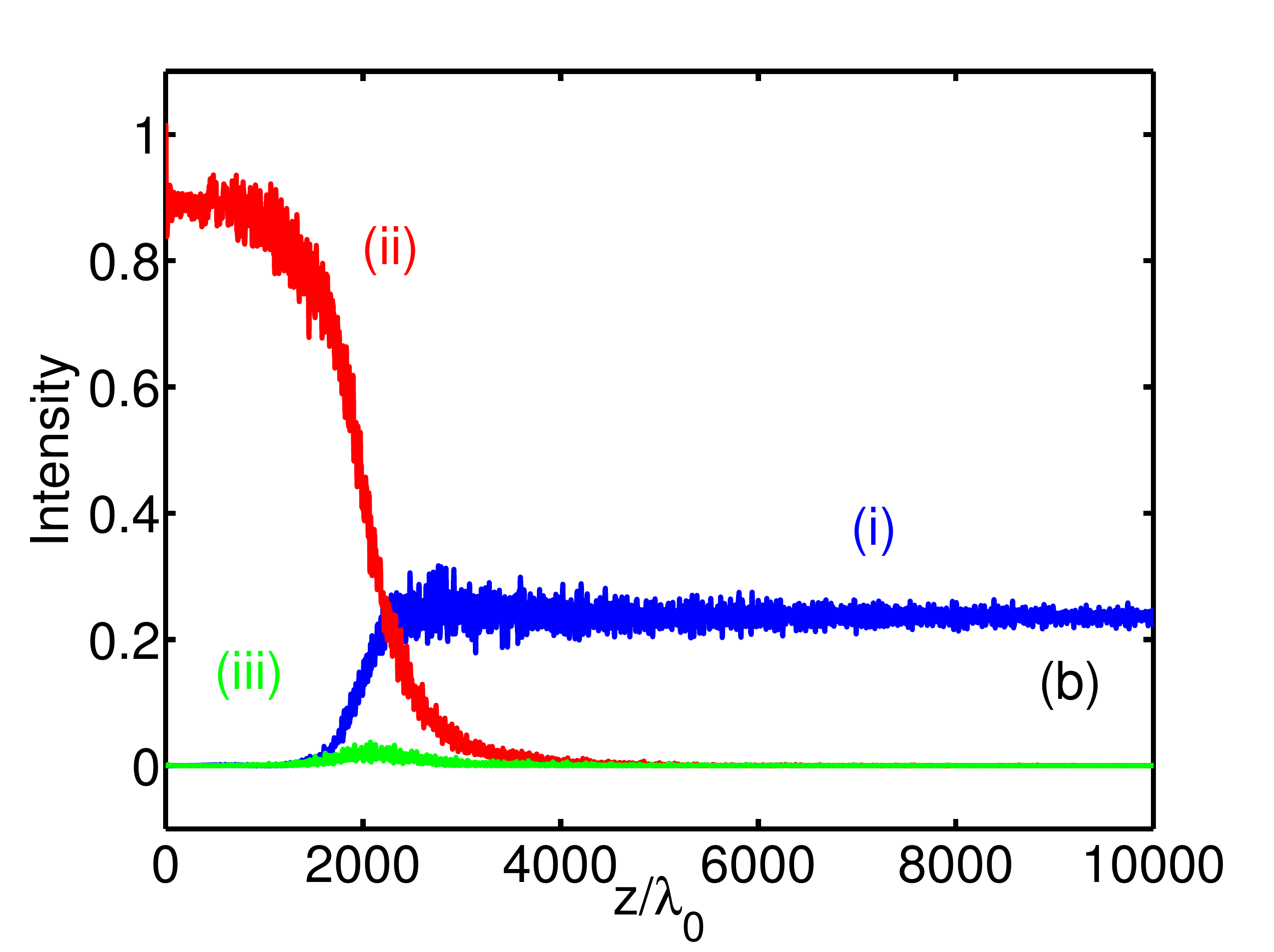}\\
 \includegraphics[width=0.45\linewidth]{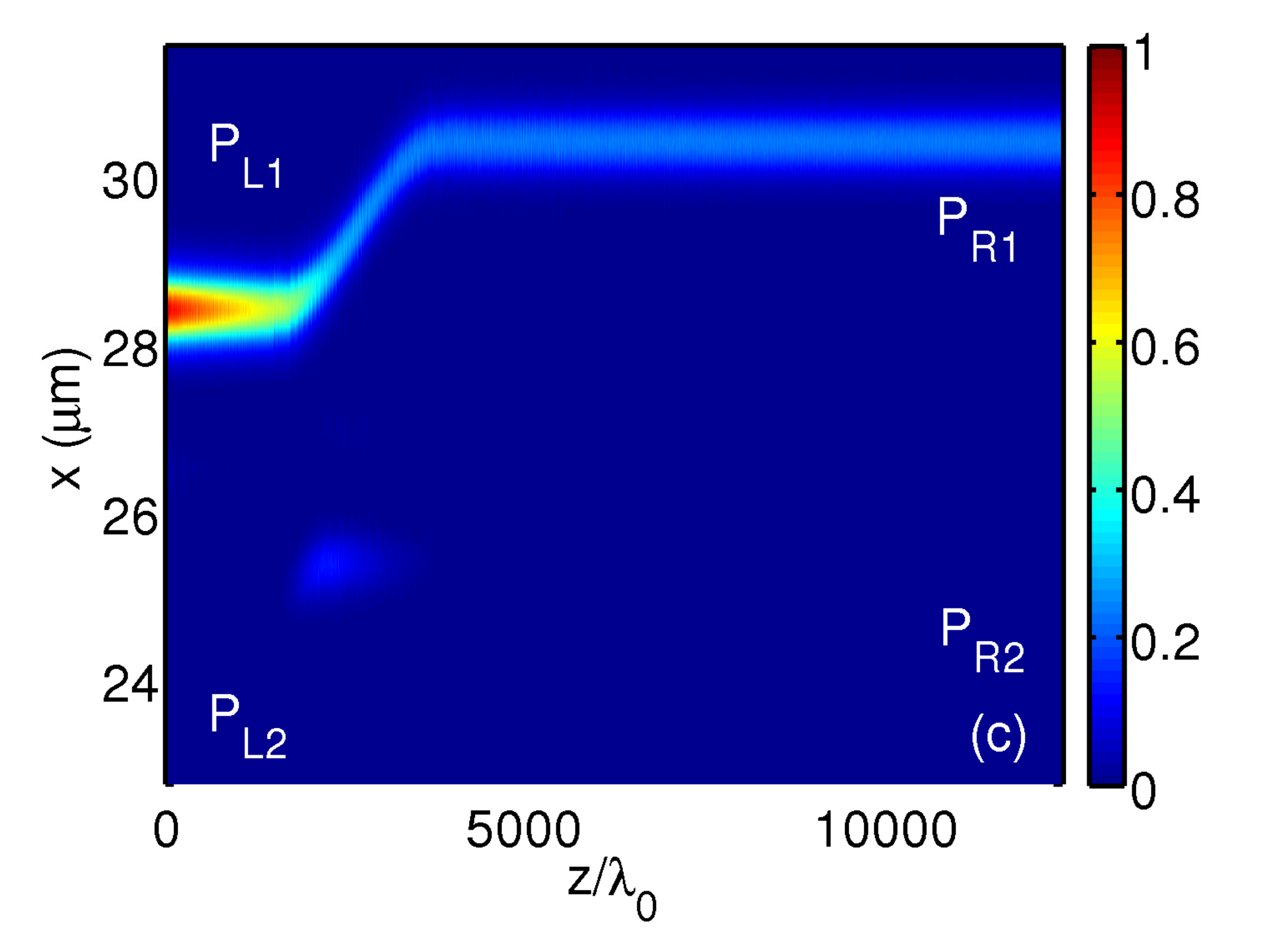}
\includegraphics[width=0.45\linewidth]{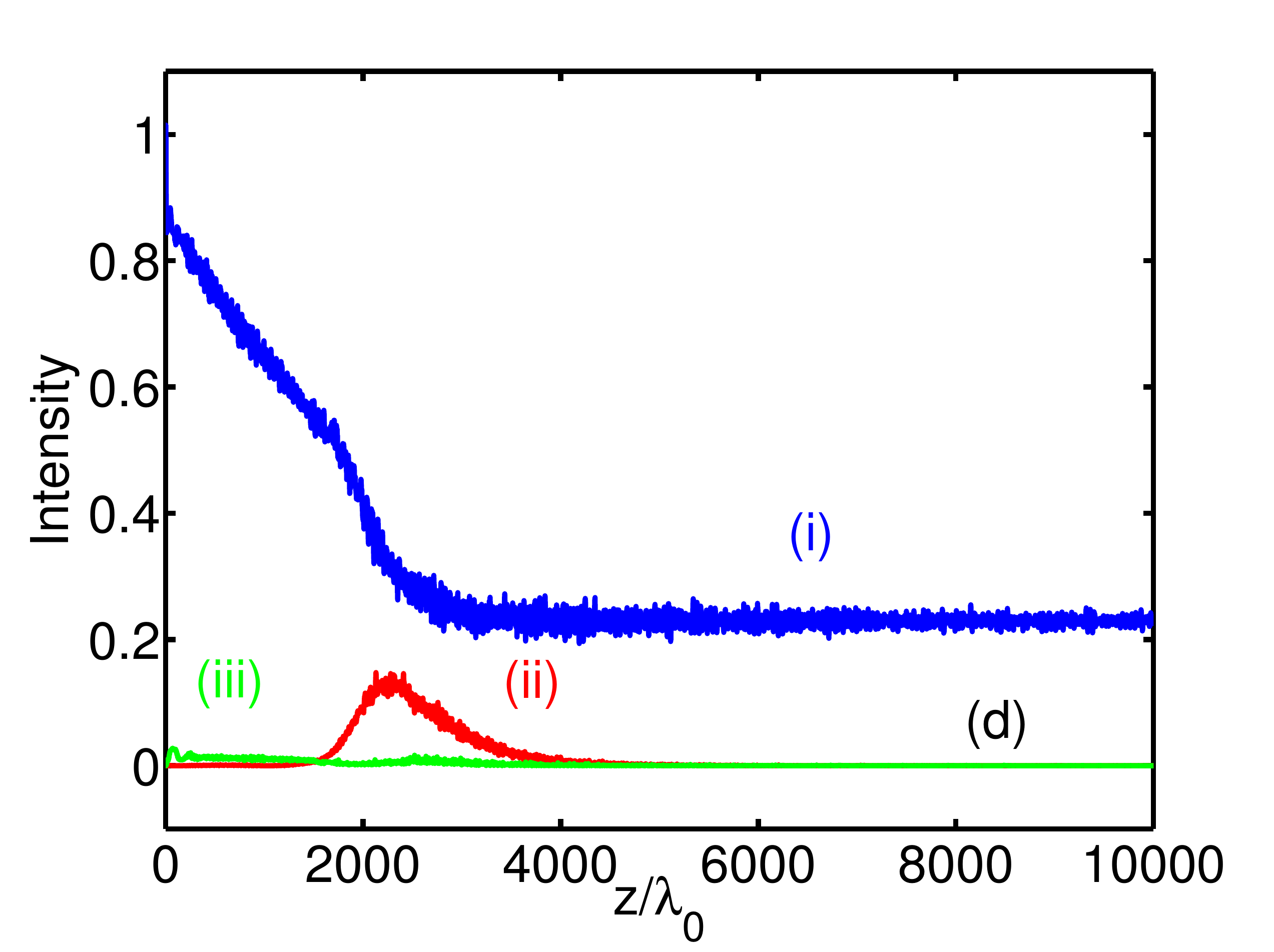}\\
\caption{\label{fig:800LRNR} Left-right nonreciprocity corresponding to Fig.~\ref{fig:800nm}(a). The field propagates from left to right in photonic circuits. (a) Light incident into the waveguide D$_2$;  (c) light enters the waveguide D$_1$; (b) and (d) Intensities of field at the middle of waveguides D$_1$ [blue lines (i)],  D$_2$ [red lines (ii)] and D$_3$ [green lines (iii)].}
\end{center}

\end{figure}

 According to Eqs.~(\ref{eq:HelmholtzE}) and ~(\ref{eq:AE}), the structure of system is scalable in size to shift the frequency window of nonrecipricity, e.g. the LRNR around $\lambda_0=800$~\nano\meter. It can be seen from the coupled mode theory Eq.~(\ref{eq:CME}) that the LRNR occurs if all parameters are scaled in a similar structure according to the propagation constant $\beta$ for a different wavelength. In Fig.~\ref{fig:800nm}(a), we first scale the photonic circuit in the x-direction and then adjust the structure parameters and the width of input light for keeping the parameters mismatching and propagation constants close to those in Fig.~\ref{fig:system}(b). As a result, the distributions of field for the inputs into ports $P_{L1}$ and $P_{L2}$ are similar to Fig.~\ref{fig:NLRNon} (see Fig.~\ref{fig:800LRNR}). There is about $25\%$ of light trapped in the waveguide D$_1$, while the light from P$_{L2}$ is vanishing small. Then the intensities outcoming from ports $P_{R1}$ and $P_{R2}$ are scanned in wavelength between $700\sim900$~\nano\meter. It can been clearly seen from Fig.~\ref{fig:800nm}(b) that the second structure allow a high performance of LRNR over $40$~\nano\meter~ from $780$~\nano\meter to $820$~\nano\meter, allowing to control a ultrashort laser pulse with duration $1/\Delta \omega \sim 30$~\femto\second. Thus our scheme promises an ultrabroadband LRNR at difference wavelengths. A shorter wavelength means a larger loss in the bending region due to the stronger dipole radiation. To reduce this unwanted loss, we use a finer grid $\delta x=50$~\nano\meter~ and $\delta z=0.5$~\micro\meter, and adjust $w_p=375$~\nano\meter~ according to the eigen mode profile in our simulation.

\section{Discussion on two  types of nonreciprocity}
Two different types of nonrecipricity, FBNR and LRNR, have been discovered in optical systems\cite{fujita2000waveguide,haldane2008possible,zheng2009observation, OE20p18440,PRL107p173902,NaturePhotonOnchip,Science335p447,NPhoton3p91,OptoAcousticIsolator1,OptoAcousticIsolator2,PRL102p213903,Optom2,hadad2010magnetized,khanikaev2010one,NPhys6p192,PRL103p093902,PRA82p043803}. The FBNR requires the breaking of the Lorentz reciprocity theorem and means that the forward and backward transmissions are not equal when the sources and responses interchange \cite{Science333p38b}. It is the basis of optical isolators. If the forward transmission $T_f$ is much larger than the backward transmission $T_{b}$, then the device with the FBNR can allow the forward propagating light to go through but block the back scattering light. Whereas the LRNR in our photonic circuit means that the light launching into different waveguides in the left hand side comes out from the same port from right. It does not violate the Lorentz reciprocity theorem \cite{Haus,RPP67p717} and as a result can not be used to isolate a light scattered backward.

In spite of the absence of ability for optical isolator, our scheme can dynamically route the light into difference paths.  
Our device also provide a novel method to switch on/off the light via dynamically tuning the loss of waveguide, shown in Fig.~\ref{fig:DEC}(c). For $\Im[\varepsilon_3]/\varepsilon_0 <-0.05$ ($\gamma \geqslant 450$ \centi\meter$^{-1}$), the light input to P$_{L2}$ can be switched off, while it can effectively transfer to the output port P$_{R1}$ for $-0.005<\Im[\varepsilon_3]/\varepsilon_0 <-0.01$. More importantly, the loss of waveguide can be tune faster ($<1$~\pico\second) and more efficiently \cite{TuneLoss1,TuneLoss2} than the refractive index modulation \cite{Nature431p1081}. Moreover, the intensity of light outcoming from port P$_{R2}$ can be increased essentially using a shorter output length.
An alternative method to route light is to dynamically tune the refractive index of waveguides.
Although our nonreciprocal photonic circuit is discussed in a linear optical medium, we also can realize the setup in a nonlinear medium like fused silica \cite{router3D} or silicon \cite{Nature431p1081} using ultrashort laser pulses. Dynamically tuning the refractive index of waveguides, see Figs.~\ref{fig:DEC}(a) and (b), and subsequently the coupling strength and phase mismatch using ultrashort laser pulses \cite{router3D}, one can switching on/off the light outcoming from port P$_{R1}$ (Fig.~\ref{fig:DEC}(a)) or route the light incident to P$_{L2}$ into port P$_{R1}$ for $10.74 \leqslant\Re[\varepsilon_3]/\varepsilon_0 \leqslant 10.78$ or port P$_{R2}$ for $\Re[\varepsilon_3]/\varepsilon_0 \leqslant 10.6$ (Fig.~\ref{fig:DEC}(b)). The dynamical tuning range is $\Delta n <1\%$, which can be obtained using the existing technology \cite{router3D,Nature431p1081,PhysRep463p1}. Therefore, our nonreciprocal design can be a ultrafast, broadband optical router.

\section{Conclusion}

In conclusion, using a technique analogous to the coherent population trapping in quantum optics, we broke the symmetry of transverse light propagation in the photonic circuits of three coupled waveguides. Our proposed system is made only from linear, passive materials. Our simulations indicate the possibility of asymmetric transverse energy flow in an ultrabroadband window spanning over $80$~nm in frequency. Although our proposed system has a relatively large insertion loss, it opens a door to the possibility of highly efficient optical nonreciprocity in a linear, passive medium.

\section*{Acknowledgments}
This research is supported by a grant from the King Abdulaziz City for Science and Technology (KACST). One of us (MSZ) is grateful for the NPRP grant 5-102-1-071 from the Qatar National Research Fund (QNRF). KX also gratefully acknowledge the hospitality at ARC Center for Engineered Quantum Systems and Department of Physics and Astronomy, Macquarie University.

\end{document}